\documentclass[11pt,oneside]{article}%
\usepackage{amssymb}
\usepackage{amsmath}
\usepackage{amsfonts}
\usepackage{graphicx}
\usepackage{cite}%
\allowdisplaybreaks
\begin{document}

\author{ Alexander Schmidt\thanks{e-mail: schmidt@theorie.physik.uni-muenchen.de},
Hartmut Wachter\thanks{e-mail: Hartmut.Wachter@physik.uni-muenchen.de}%
\vspace{0.16in}\\Max-Planck-Institute\\for Mathematics in the Sciences\\Inselstr. 22, D-04103 Leipzig, Germany\vspace{0.16in}\\Arnold-Sommerfeld-Center\\Ludwig-Maximilians-Universit\"{a}t\\Theresienstr. 37, D-80333 M\"{u}nchen, Germany}
\title{q-Deformed quantum Lie algebras}
\maketitle
\date{}

\begin{abstract}
\noindent Attention is focused on q-deformed quantum algebras with physical
importance, i.e. $U_{q}(su_{2})$, $U_{q}(so_{4})$ and q-deformed Lorentz
algebra. The main concern of this article is to assemble important ideas about
these symmetry algebras in a consistent framework which shall serve as
starting point for representation theoretic investigations in physics,
especially quantum field theory. In each case considerations start from a
realization of symmetry generators within the differential algebra. Formulae
for coproducts and antipodes on symmetry generators are listed. The action of
symmetry generators in terms of their Hopf structure is taken as q-analog of
classical commutators and written out explicitly. Spinor and vector
representations of symmetry generators are calculated. A review of the
commutation relations between symmetry generators and components of a spinor
or vector operator is given. Relations for the corresponding quantum Lie
algebras are computed. Their Casimir operators are written down in a form
similar to the undeformed case.\newpage

\end{abstract}

\section{Introduction}

It is an old idea that limiting the precision of position measurements by a
fundamental length will lead to a new method for regularizing quantum field
theories \cite{Heis38} . It is also well-known that such a modification of
classical spacetime will in general break its Poincar\'{e} symmetry
\cite{Sny47}. One way out of this difficulty is to change not only spacetime,
but also its underlying symmetry.

Quantum groups can be seen as deformations of classical spacetime symmetries,
as they describe the symmetry of their comodules, the so-called quantum
spaces. From a physical point of view the most realistic examples for\ quantum
groups and quantum spaces arise from q-deformation \cite{Ku83, Dri85, Drin86,
Jim85, Wor87, Man88, RFT90}. In our work we are interested in q-deformed
versions of Minkowski space and Euclidean spaces as well as their
corresponding symmetries, given by q-deformed Lorentz algebra and algebras of
q-deformed angular momentum, respectively\ \cite{CSSW90, Pod90, SWZ91, Maj91,
LWW97}. Remarkably, Julius Wess and his coworkers were able to show that
q-deformation of spaces and symmetries can indeed lead to discretizations, as
they result from the existence of a smallest distance \cite{Fich97, CW98}.
This observation nourishes the hope that q-deformation might give a new method
to regularize quantum field theories \cite{MajReg, GKP96, Oec99, Blo03}.

In our previous work \cite{WW01, BW01, Wac02, Wac04,WacTr, Mik04,SW04}
attention was focused on q-deformed quantum spaces of physical importance,
i.e. two-dimensional Manin plane, q-deformed Euclidean space in three or four
dimensions and q-deformed Minkowski space. If we want to describe fields on
q-deformed quantum spaces we need to consider representations of the
corresponding quantum symmetries, given by $U_{q}(su_{2})$, $U_{q}(so_{4})$
and q-deformed Lorentz algebra. The study of such quantum algebras has
produced a number of remarkable results during the last two decades. For a
review we recommend the reader the presentations in \cite{KS97, Maj95, ChDe96}
and references therein. In this article we want to adapt these general ideas
to our previous considerations about q-deformed quantum spaces. In doing so,
we provide a basis for performing concrete calculations, as they are necessary
in formulating and evaluating field theories on quantum spaces.

In particular, we intend to proceed as follows. In Sec. \ref{BasSec} we cover
the ideas our considerations about q-deformed quantum symmetries are based on.
In the subsequent sections we first recall for each quantum algebra under
consideration how its generators are realized within the corresponding
q-deformed\ differential calculus. Then we are going to present explicit
formulae for coproduct and antipode on a set of independent symmetry
generators. With this knowledge at hand we should be able to write down
explicit formulae for so-called q-commutators between symmetry generators and
representation space elements. In addition to this, we are going to consider
spinor and vector representations of the independent symmetry generators and
give a complete review of the commutation relations between symmetry
generators and components of a spinor or vector operator. Furthermore we are
going to calculate the adjoint action of the independent symmetry generators
on each other. In this manner, we will get relations for quantum Lie algebras.
We will close our considerations by writing down q-analogs of Casimir
operators. Finally, Sec. \ref{Concl} shall serve as a short conclusion. For
reference and for the purpose of introducing consistent and convenient
notation, we provide a review of key notations and results in Appendix
\ref{AppA}.

We should also mention that most of our results were obtained by applying the
computer algebra system Mathematica \cite{Wol}. We are convinced that in the
future this powerful tool will be inevitable in managing the extraordinary
complexity of q-deformation.

\section{Basic ideas on q-deformed quantum symmetries\label{BasSec}}

Roughly speaking a quantum space is nothing else than an algebra generated by
non-commuting coordinates $X_{1},X_{2},\ldots,X_{n},$ i.e.%
\begin{equation}
\mathcal{A}_{q}=\mathbb{C}{[}{[}X_{1},\ldots X_{n}{]}{]}/\mathcal{I},
\end{equation}
where $\mathcal{I}$ denotes the ideal generated by the relations of the
non-commuting coordinates. The quantum spaces we are interested in for
physical reasons are two-dimensional Manin plane, q-deformed Euclidean space
in three or four dimensions as well as q-deformed Minkowski space. For their
definition we refer the reader to Appendix \ref{AppA}.

On each of these quantum spaces exist two differential calculi
\cite{WZ91,CSW91,Song92} with
\begin{align}
\partial^{i}X^{j}  &  =g^{ij}+k(\hat{R}^{-1})_{kl}^{ij}X^{k}\partial
^{\,l},\qquad k\in\mathbb{R},\\
\hat{\partial}^{i}X^{j}  &  =g^{ij}+k^{-1}(\hat{R})_{kl}^{ij}X^{k}%
\hat{\partial}^{\,l},\nonumber
\end{align}
where $\hat{R}$ and $g^{ij}$ denote respectively the $R$-matrix and the
corresponding quantum metric of the underlying quantum symmetry. There is a
q-deformed antisymmetriser $P_{A}$ that enables us to define the components of
q-deformed orbital angular momentum as antisymmetrized products of coordinates
and derivatives \cite{Fio95}. Specifically, we have%
\begin{equation}
L^{ij}\sim(P_{A})_{kl}^{ij}\,X^{k}\partial^{\,l}\sim(P_{A})_{kl}^{ij}%
\,X^{k}\hat{\partial}^{\,l}.
\end{equation}
It can be shown that the $L^{ij}$ together with a central generator $U$ being
subject to
\begin{equation}
U-1\sim\lambda g_{ij}\,X^{i}\hat{\partial}^{\,j},\quad\lambda\equiv q-q^{-1},
\end{equation}
span the quantum algebras describing the underlying quantum symmetry. This way
we will obtain $U_{q}(su_{2})$, $U_{q}(so_{4})$ as well as q-deformed Lorentz
algebra in the sequel of this article.

In accordance with the classical case, a single particle wave function should
be defined as element of the tensor product of a finite vector space
$\mathcal{S}$ holding the spin degrees of freedom and the algebra of space
functions $\mathcal{M}$, where $\mathcal{S}$ and $\mathcal{M}$ are both
modules of the q-deformed symmetry algebra under consideration. In contrast to
the classical situation, we now have to distinguish between right and left
wave functions, i.e.
\begin{equation}
\psi_{R}\equiv\sum\nolimits_{j}e^{j}\otimes\psi^{j},\quad\psi_{L}\equiv
\sum\nolimits_{j}\psi^{j}\otimes e^{j},
\end{equation}
where ${\{}e^{j}{\}}$ is a basis in $\mathcal{S}$ and $\psi^{j}$ are elements
of $\mathcal{M}$. The reason for this lies in the fact that due to the
braiding between the two tensor factors we are not allowed to interchange them
by a trivial flip \cite{Maj94Kat}.

The transformation properties of our\ wave functions are determined by the
coproduct of the symmetry generators, since we have (we write the coproduct in
the so-called Sweedler notation, i.e. $\Delta(a)=a_{(1)}\otimes a_{(2)})$
\begin{align}
L^{ij}\rhd\psi_{R}  &  =\sum\nolimits_{k}(L^{ij})_{(1)}\rhd e^{k}%
\otimes(L^{ij})_{(2)}\rhd\psi^{k},\\
L^{ij}\rhd\psi_{L}  &  =\sum\nolimits_{k}(L^{ij})_{(1)}\rhd\psi^{k}%
\otimes(L^{ij})_{(2)}\rhd e^{k}.\nonumber
\end{align}
For scalar wave functions $\mathcal{S}$ has to be a one-dimensional vector
space with the corresponding representation on it being the trivial one. In
this manner, the symmetry transformation is reduced to the action on space
functions. In Ref. \cite{BW01} we derived explicit formulae for such actions.
In the sequel of this article we will restrict ourselves to representations of
symmetry generators on a spinorial or vectorial basis. Physical fields of
higher spin should then be built up from spinor or vector fields.

It should be noted that we can combine a quantum algebra $\mathcal{H}$ with
its representation space $\mathcal{A}$ to form a left cross product algebra
$\mathcal{A}\rtimes\mathcal{H}$ built on $\mathcal{A}\otimes\mathcal{H}$ with
product%
\begin{equation}
(a\otimes h)(b\otimes g)=a(h_{(1)}\triangleright b)\otimes h_{(2)}g,\quad
a,b\in\mathcal{A},\mathcal{\quad}h,g\in\mathcal{H}. \label{LefCrosPro}%
\end{equation}
There is also a right-handed version of this notion called a right cross
product algebra $\mathcal{H}\ltimes\mathcal{A}$ and built on $\mathcal{H}%
\otimes\mathcal{A}$ with product%
\begin{equation}
(h\otimes a)(g\otimes b)=hg_{(1)}\otimes(a\triangleleft g_{(2)})b.
\label{RigCrosPro}%
\end{equation}
The last two identities tell us that the commutation relations between
symmetry generators and representation space elements are completely
determined by coproduct and action of the symmetry generators.

In this article it can be seen that there is a remarkable correspondence
between q-deformed symmetry algebras and their classical counterparts. Towards
this end we have to introduce the notion of a q-commutator which is nothing
other than the action of a symmetry generator $L^{ij}$ on a representation
space element $V.$ Expressing that action in terms of the Hopf structure of
$L^{ij}$ the q-brackets become%
\begin{align}
{[}L^{ij},V{]}_{q}  &  \equiv L^{ij}\triangleright V=L_{(1)}^{ij}%
VS(L_{(2)}^{ij}),\label{q-KommAllg}\\
{[}V,L^{ij}{]}_{q}  &  \equiv V\triangleleft L^{ij}=S^{-1}(L_{(2)}%
^{ij})VL_{(1)}^{ij},\nonumber
\end{align}
where $S$ and $S^{-1}$ denote the antipode and its inverse, respectively. From
their very definition it follows that q-commutators obey the q-deformed Jacobi
identities
\begin{align}
{[}L^{ij},{[}L^{kl},V{]}_{q}{]}_{q}  &  ={[}{[}L_{(1)}^{ij},L^{kl}{]}_{q}%
,{[}L_{(2)}^{ij},V{]}_{q}{]}_{q},\\
{[}{[}L^{ij},L^{kl}{]}_{q},V{]}_{q}  &  ={[}L_{(1)}^{ij},{[}L^{kl}%
,{[}S(L_{(2)}^{ij}),V{]}_{q}{]}_{q}{]}_{q}.\nonumber
\end{align}

Now, we are able to introduce the notion of a quantum Lie algebra as it was
given in \cite{Schupp93, Maj94, Sud92}. A quantum Lie algebra can be regarded
as a subspace of a q-deformed enveloping algebra $U_{q}(g)$ being invariant
under the adjoint action of $U_{q}(g)$. The point now is that the $L^{ij}$ are
the components of a tensor operator and this is the reason why their adjoint
action on each other equals to a linear combination of the $L^{ij}$
\cite{Bie90}, i.e. the $L^{ij}$ span a quantum Lie algebra with
\begin{equation}
{[}L^{ij},L^{kl}{]}_{q}(=L^{ij}\triangleright L^{kl}=L^{ij}\triangleleft
L^{kl})=(C^{ij})_{mn}^{kl}L^{mn}, \label{Liecom}%
\end{equation}
where the $C^{ij}$ are the so-called quantum structure constants. In the
subsequent sections we are going to determine those constants in the case of
$U_{q}(su_{2})$, $U_{q}(so_{4})$ and q-deformed Lorentz algebra.

Finally, let us mention that the q-brackets give another way to write down
commutation relations between symmetry generators and components of a\ vector
or spinor operator. More formally, we have
\begin{equation}
{[}L^{ij},X^{k}{]}_{q}=(\tau_{L}^{ij})^{k}{}_{l}\,X^{l},\quad{[}X_{k}%
,L^{ij}{]}_{q}=X_{l}\,(\tau_{R}^{ij})^{l}{}_{k}, \label{VerLX}%
\end{equation}
and
\begin{equation}
{[}L^{ij},\theta^{\alpha}{]}_{q}=(\sigma_{L}^{ij})^{\alpha}{}_{\beta}%
\,\theta^{\beta},\quad\left[  \theta^{\alpha},L^{ij}\right]  _{q}%
=\theta_{\beta}\,(\sigma_{R}^{ij})^{\beta}{}_{\alpha}, \label{VerLTheta}%
\end{equation}
where $X^{k}$ and $\theta^{\alpha}$ stand respectively for components of
vector and spinor operators. Notice that for $\tau_{L/R}$ and $\sigma_{L/R}$
we have to substitute the representation matrix of $L^{ij}$, as it comes out
for the vector and spinor representation, respectively. It should be
appreciated that the relations in (\ref{Liecom}), (\ref{VerLX}), and
(\ref{VerLTheta}) become part of a q-deformed Super-Euclidean or
Super-Poincar\'{e} algebra, if such objects exist.

\section{Quantum Lie algebra of three-dimensional angular
momentum\label{QuLie3dim}}

\subsection{Representation of three-dimensional angular momentum within
q-deformed differential calculus}

In the case of three-dimensional $q$-deformed Euclidean space (for its
definition see Appendix \ref{AppA}) the generators of orbital angular momentum
are defined by \cite{LWW97}
\begin{equation}
L^{A}\equiv\Lambda^{1/2}X^{C}\hat{\partial}^{D}\epsilon_{DC}{}^{A},\quad
A\in\{+,3,-\}, \label{DrehGen3dim}%
\end{equation}
where $\varepsilon_{DC}{}^{A}$ denotes a q-analog of the completely
antisymmetric tensor of third rank and $\Lambda$ stands for a scaling operator
subject to
\begin{equation}
\Lambda X^{A}=q^{4}X^{A}\Lambda,\quad\Lambda\hat{\partial}^{A}=q^{-4}%
\hat{\partial}^{A}\Lambda,\quad A\in\{+,3,-\}.
\end{equation}
If not stated otherwise summation over all repeated indices is to be
understood. Substituting for\ $\varepsilon_{DC}{}^{A}$ their explicit form, we
obtain from Eq. (\ref{DrehGen3dim})
\begin{align}
L^{+}  &  =-q^{-1}\Lambda^{1/2}X^{+}\hat{\partial}^{3}+q^{-3}\Lambda
^{1/2}X^{3}\hat{\partial}^{+},\\
L^{3}  &  =-q^{-2}\Lambda^{1/2}X^{+}\hat{\partial}^{-}+q^{-2}\Lambda
^{1/2}X^{-}\hat{\partial}^{+}-q^{-2}\lambda\Lambda^{1/2}X^{3}\hat{\partial
}^{3},\nonumber\\
L^{-}  &  =-q^{-1}\Lambda^{1/2}X^{3}\hat{\partial}^{-}+q^{-3}\Lambda
^{1/2}X^{-}\hat{\partial}^{3}.\nonumber
\end{align}

Using the Leibniz rules for partial derivatives in the form
\begin{equation}
X^{A}\hat{\partial}^{B}=g^{AB}+(\hat{R}^{-1})^{AB}{}_{CD}\,\hat{\partial}%
^{C}X^{D},
\end{equation}
and taking into account the identities%
\begin{equation}
g^{AB}\epsilon_{BA}{}^{C}=0,\quad(\hat{R}^{-1})^{AB}{}_{CD}\,\epsilon_{BA}%
{}^{E}=-q^{4}\epsilon_{DC}{}^{E},
\end{equation}
the generators in Eq. (\ref{DrehGen3dim}) can alternatively be written as
\begin{equation}
L^{A}=-q^{4}\Lambda^{1/2}\hat{\partial}^{C}X^{D}\epsilon_{DC}{}^{A}.
\end{equation}
As already mentioned, there is a second set of derivatives ${\partial}^{A}$
which can be linked to the first one via the relation \cite{LWW97}%
\begin{equation}
\hat{\partial}^{A}=\Lambda^{-1}(\partial^{A}+q^{3}\lambda X^{A}\partial
^{B}\partial^{C}g_{BC}).
\end{equation}
Making use of this identity together with $X^{C}X^{D}\epsilon_{DC}{}^{A}=0$,
one can show that we additionally have
\begin{equation}
L^{A}=q^{4}\Lambda^{-1/2}X^{C}\partial^{D}\epsilon_{DC}{}^{A}=-\Lambda
^{-1/2}\partial^{C}X^{D}\epsilon_{DC}{}^{A}.
\end{equation}

\subsection{Hopf structure of $U_{q}(su_{2})$ and corresponding q-commutators}

As it is well-known, the generators $L^{+},$ $L^{3},$ and $L^{-}$ together
with a generator $\tau^{1/2}$ can be viewed as elements of the quantum algebra
$U_{q}(su_{2})$ \cite{LWW97}. Its defining relations read
\begin{gather}
\tau^{1/2}L^{\pm}=q^{\mp2}L^{\pm}\tau^{1/2},\quad\tau^{1/2}L^{3}=L^{3}%
\tau^{1/2},\label{UqSu2}\\
L^{-}L^{+}-L^{+}L^{-}=\tau^{-1/2}L^{3},\nonumber\\
L^{\pm}L^{3}-L^{3}L^{\pm}=\mp q^{\pm1}L^{\pm}\tau^{-1/2}.\nonumber
\end{gather}
Furthermore, this algebra has a Hopf structure given by
\begin{align}
\Delta(L^{\pm})  &  =L^{\pm}\otimes\tau^{-1/2}+1\otimes L^{\pm},\\
\Delta(L^{3})  &  =L^{3}\otimes\tau^{-1/2}+\tau^{1/2}\otimes{L^{3}}\nonumber\\
&  +\;\lambda\tau^{1/2}(qL^{+}\otimes L^{-}+q^{-1}L^{-}\otimes{L^{+}%
}),\nonumber\\[0.16in]
S(L^{\pm})  &  =q^{\mp2}S^{-1}(L^{\pm})=-L^{\pm}\tau^{1/2},\\
S(L^{3})  &  =S^{-1}(L^{3})=-q^{-2}L^{3}+\lambda\lambda_{+}\tau^{1/2}%
L^{+}L^{-},\nonumber\\[0.16in]
\varepsilon(L^{A})  &  =0,\quad A\in\{+,3,-\},
\end{align}
where $\lambda_{+}\equiv q+q^{-1}$. In addition to this, let us notice that
$\tau^{1/2}$ is a grouplike generator.

With this Hopf structure at hand we are in a position to write down
expressions for\ q-commutators. Specifically, we get from Eq.
(\ref{q-KommAllg}) for left-commutators
\begin{align}
{[}L^{\pm},V{]}_{q}  &  =(L^{\pm}V-VL^{\pm})\tau^{1/2},\label{commeu3l}\\
{[}L^{3},V{]}_{q}  &  =L^{3}V\tau^{1/2}-q^{-2}\tau^{1/2}VL^{3}\nonumber\\
&  -\;\lambda\tau^{1/2}(q^{-1}L^{+}V\tau^{1/2}L^{-}+qL^{-}V\tau^{1/2}%
L^{+})\nonumber\\
&  +\;\lambda\lambda_{+}\tau^{1/2}V\tau^{1/2}L^{+}L^{-},\nonumber
\end{align}
and likewise for\ right-commutators,%
\begin{align}
{[}V,L^{\pm}{]}_{q}  &  =\tau^{1/2}(VL^{\pm}-L^{\pm}V)\label{comeu3r}\\
{[}V,L^{3}{]}_{q}  &  =\tau^{1/2}VL^{3}-q^{-2}L^{3}V\tau^{1/2}\nonumber\\
&  -\;\lambda\tau^{1/2}(q^{-1}L^{+}V\tau^{1/2}L^{-}+qL^{-}V\tau^{1/2}%
L^{+})\nonumber\\
&  +\;\lambda\lambda_{+}\tau^{1/2}L^{+}L^{-}V\tau^{1/2},\nonumber
\end{align}
where $V$ denotes an element living in a representation space of $U_{q}%
(su_{2}).$

\subsection{Matrix representations of $U_{q}(su_{2})$ and commutation
relations with tensor operators}

Next, we would like to turn our attention to some special representations of
the symmetry generators $L^{+},L^{3},$ and $L^{-}$, namely spinor and vector
representations \cite{KS97,Bie95}. The finite dimensional representations of
$U_{q}(su_{2})$ are already well-known (see for example Refs. \cite{KS97} and
\cite{Bie95}). With our conventions (see also Appendix \ref{AppA}) the spinor
representations on symmetry generators become%
\begin{align}
L^{A}\rhd\theta^{\,\alpha}  &  =(\sigma^{A})^{\alpha}{}_{\beta}\;\theta
^{\beta}, & \tau^{1/2}\rhd\theta^{\,\alpha}  &  =(\tau^{1/2})^{\alpha}%
{}_{\beta}\;\theta^{\beta},\label{RepDreh2dim}\\
\theta_{\alpha}\lhd L^{A}  &  =\theta_{\beta}\,(\sigma^{A})^{\beta}{}_{\alpha
}, & \theta_{\alpha}\lhd\tau^{1/2}  &  =\theta_{\beta}\,(\tau^{1/2})^{\beta}%
{}_{\alpha},\nonumber
\end{align}
where we have introduced as some kind of q-deformed sigma matrices
\begin{gather}
(\sigma^{+})^{\alpha}{}_{\beta}=-q^{1/2}\lambda_{+}^{-1/2}\left(
\begin{array}
[c]{cc}%
0 & 0\\
1 & 0
\end{array}
\right)  ,\quad(\sigma^{-})^{\alpha}{}_{\beta}=q^{-1/2}\lambda_{+}%
^{-1/2}\left(
\begin{array}
[c]{cc}%
0 & 1\\
0 & 0
\end{array}
\right)  ,\label{SigmMatr3dim}\\
(\sigma^{3})^{\alpha}{}_{\beta}=\lambda_{+}^{-1}\left(
\begin{array}
[c]{cc}%
-q & 0\\
0 & q^{-1}%
\end{array}
\right)  ,\quad(\tau^{1/2})^{\alpha}{}_{\beta}=\left(
\begin{array}
[c]{cc}%
q & 0\\
0 & q^{-1}%
\end{array}
\right)  .\nonumber
\end{gather}
Here and in what follows, we shall take the convention that \textbf{upper and
lower matrix indices refer to columns and rows, respectively}. That the above
matrices indeed give a representation can easily be checked. Towards this end,
we have to substitute in (\ref{UqSu2}) the sigma matrices for the algebra
generators. Then we can show by usual matrix multiplication that the\ algebra
relations are fulfilled.

The above results enable us to write down commutation relations between
symmetry generators and components of a spinor operator. Such a spinor
operator with components $\theta^{\,\alpha},$ $\alpha=1,2,$ is completely
determined by its transformation properties%
\begin{equation}
{[}L^{A},\theta^{\,\alpha}{]}_{q}=L^{A}\rhd\theta^{\,\alpha},\quad{[}%
\theta_{\alpha},L^{A}{]}_{q}=\theta_{\alpha}\lhd L^{A}. \label{TransTensOp}%
\end{equation}
Inserting the results of (\ref{RepDreh2dim}) and (\ref{SigmMatr3dim}) into the
relations of (\ref{TransTensOp}) and then rearranging, it follows that
\begin{align}
L^{+}\theta^{1}  &  =\theta^{1}L^{+}-q^{1/2}\lambda_{+}^{-1/2}\theta^{2}%
\tau^{-1/2},\\
L^{+}\theta^{2}  &  =\theta^{2}L^{+},\nonumber\\
L^{3}\theta^{1}  &  =q\theta^{1}L^{3}-q^{-1/2}\lambda\lambda_{+}^{-1/2}%
\theta^{2}L^{-}-q\lambda_{+}^{-1}\theta^{1}\tau^{-1/2},\nonumber\\
L^{3}\theta^{2}  &  =q^{-1}\theta^{2}L^{3}+q^{1/2}\lambda\lambda_{+}%
^{-1/2}\theta^{1}L^{+}+q^{-1}\lambda_{+}^{-1}\theta^{2}\tau^{-1/2},\nonumber\\
L^{-}\theta^{1}  &  =\theta^{1}L^{-},\nonumber\\
L^{-}\theta^{2}  &  =\theta^{2}L^{-}+q^{-1/2}\lambda_{+}^{-1/2}\theta^{1}%
\tau^{-1/2},\nonumber
\end{align}
and likewise,
\begin{align}
\theta_{1}L^{+}  &  =L^{+}\theta_{1},\\
\theta_{2}L^{+}  &  =L^{+}\theta_{2}-q^{1/2}\lambda_{+}^{-1/2}\tau
^{-1/2}\theta_{1},\nonumber\\
\theta_{1}L^{3}  &  =qL^{3}\theta_{1}+q^{-1/2}\lambda\lambda_{+}^{-1/2}%
L^{+}\theta_{2}+q\lambda_{+}^{-1}\tau^{-1/2}\theta_{1},\nonumber\\
\theta_{2}L^{3}  &  =q^{-1}L^{3}\theta_{2}-q^{1/2}\lambda\lambda_{+}%
^{-1/2}L^{-}\theta_{1}+q^{-1}\lambda_{+}^{-1}\tau^{-1/2}\theta_{2},\nonumber\\
\theta_{1}L^{-}  &  =L^{-}\theta_{1}+q^{-1/2}\lambda_{+}^{-1/2}\tau
^{-1/2}\theta_{2},\nonumber\\
\theta_{2}L^{-}  &  =L^{-}\theta_{2}.\nonumber
\end{align}

Next, we turn to the vector representations of $U_{q}(su_{2})$. They take the
form
\begin{equation}
L^{A}\rhd X^{B}=(\tau^{A})^{B}{}_{C}\;X^{C},\quad X_{B}\lhd L^{A}=X_{C}%
\,(\tau^{A})^{C}{}_{B},
\end{equation}
with
\begin{gather}
(\tau^{+})^{B}{}_{C}=\left(
\begin{array}
[c]{ccc}%
0 & -q & 0\\
0 & 0 & -1\\
0 & 0 & 0
\end{array}
\right)  ,\quad(\tau^{-})^{B}{}_{C}=\left(
\begin{array}
[c]{ccc}%
0 & 0 & 0\\
-1 & 0 & 0\\
0 & q^{-1} & 0
\end{array}
\right)  ,\\
(\tau^{3})^{B}{}_{C}=\left(
\begin{array}
[c]{ccc}%
q^{-1} & 0 & 0\\
0 & -\lambda & 0\\
0 & 0 & -q
\end{array}
\right)  ,\quad(\tau^{1/2})^{B}{}_{C}=\left(
\begin{array}
[c]{ccc}%
q^{-2} & 0 & 0\\
0 & 1 & 0\\
0 & 0 & q^{2}%
\end{array}
\right)  ,\nonumber
\end{gather}
where rows and columns are arranged in the order $+,$ $3,$ and $-$(from left
to right and top to bottom). Their property of being a representation can be
verified in very much the same way as was done for the spinor representation.
Finally, let us note that the above representation matrices and the q-deformed
$\varepsilon$-tensor are linked via
\begin{equation}
(\tau^{A})^{B}{}_{C}=q^{2}\varepsilon^{AB}{}_{C},\quad A\in\{+,3,-\}.
\end{equation}

With the same reasonings already applied to the case of spinor representations
we can derive commutation relations between symmetry generators and the
components of a vector operator. In complete accordance to the spinor case,
their general form is given by%
\begin{equation}
{[}L^{A},X^{B}{]}_{q}=L^{A}\rhd X^{B},\quad{[}X_{B},L^{A}{]}_{q}=X_{B}\lhd
L^{A}, \label{VecDar3dim}%
\end{equation}
from which we find by specifying q-commutator and vector representation%
\begin{align}
L^{\pm}X^{\pm}  &  =X^{\pm}L^{\pm},\\
L^{\pm}X^{\mp}  &  =X^{\mp}L^{\pm}\mp X^{3}\tau^{-1/2},\nonumber\\
L^{\pm}X^{3}  &  =X^{3}L^{\pm}\mp q^{\pm1}X^{\pm}\tau^{-1/2},\nonumber\\
L^{3}X^{\pm}  &  =q^{\mp2}X^{\pm}L^{3}\pm q^{\mp1}\lambda X^{3}L^{\pm}\pm
q^{\mp1}X^{\pm}\tau^{-1/2},\nonumber\\
L^{3}X^{3}  &  =X^{3}L^{3}+\lambda(X^{-}L^{+}-X^{+}L^{-})-\lambda X^{3}%
\tau^{-1/2},\nonumber
\end{align}
and
\begin{align}
X_{\mp}L^{\pm}  &  =L^{\pm}X_{\mp},\\
X_{\pm}L^{\pm}  &  =L^{\pm}X_{\pm}\mp q^{\pm1}\tau^{-1/2}X_{3},\nonumber\\
X_{3}L^{\pm}  &  =L^{\pm}X_{3}\mp\tau^{-1/2}X_{\mp},\nonumber\\
X_{\pm}L^{3}  &  =q^{\mp2}L^{3}X_{\pm}\mp\lambda L^{\mp}X^{3}\pm q^{\mp1}%
\tau^{-1/2}X_{\pm},\nonumber\\
X_{3}L^{3}  &  =L^{3}X_{3}+\lambda(q^{-1}L^{+}X^{+}-qL^{-}X_{-})-\lambda
\tau^{-1/2}X_{3}.\nonumber
\end{align}
This way, we regain the commutation relations already presented in
\cite{LWW97}.

\subsection{Quantum Lie algebra of $U_{q}(su_{2})$ and its Casimir operator}

Recalling that $L^{A}$, $A\in\left\{  +,3,-\right\}  ,$ is a vector operator,
the identities in (\ref{VecDar3dim}) also apply to the case with $X^{A}$ being
replaced by $L^{A}$. In doing so, we are led to the relations for the quantum
Lie algebra of $U_{q}(su_{2})$, i.e.
\begin{equation}
{[}L^{A},L^{B}{]}_{q}=q^{2}\varepsilon^{AB}{}_{C}\,L^{C},
\end{equation}
or more concretely,%
\begin{align}
{[}L^{A},L^{A}{]}_{q}  &  =0,\qquad A\in\left\{  +,-\right\}  ,\\
{[}L^{3},L^{3}{]}_{q}  &  =-\lambda L^{3},\nonumber\\
{[}L^{\pm},L^{3}{]}_{q}  &  =\mp q^{\pm1}L^{\pm},\nonumber\\
{[}L^{\pm},L^{\mp}{]}_{q}  &  =\mp L^{3}.\nonumber
\end{align}
These relations are equivalent to those in (\ref{UqSu2}), as can be seen in a
straightforward manner by writing out the q-commutators explicitly.

Instead of the vector operator $L^{A}$, one can just as well use a complete
antisymmetric tensor operator of second rank given by
\begin{equation}
M^{AB}\equiv\varepsilon^{AB}{}_{C}\,L^{C}.
\end{equation}
Its antisymmetry requires the following identities to hold, which can easily
be read off from its very definition:%
\begin{gather}
M^{++}=M^{--}=0,\quad M^{\pm3}=-q^{\pm2}M^{3\pm},\\
M^{-+}=-M^{+-},\quad M^{33}=\lambda M^{+-}.\nonumber
\end{gather}
Consequently, we have three independent components, for which we can choose%
\begin{equation}
M^{3\pm}=-q^{\pm1}L^{\pm},\quad M^{+-}=-q^{-2}L^{3}.
\end{equation}

Last but not least we want to deal with the Casimir operator of our quantum
Lie algebra. Let us recall that a Casimir operator $C$ has to be subject to
\begin{equation}
{[}L^{A},C{]}_{q}={[}C,L^{A}{]}_{q}=0\quad\mbox{for all}\quad A\in\left\{
+,3,-\right\}  . \label{CasOp}%
\end{equation}
By a direct calculation making use of
\begin{align}
{[}L^{A},UV{]}_{q}  &  ={[}L_{(1)}^{A},U{]}_{q}{[}L_{(2)}^{A},V{]}_{q},\\
{[}UV,L^{A}{]}_{q}  &  ={[}U,L_{(2)}^{A}{]}_{q}{[}V,L_{(1)}^{A}{]}%
_{q},\nonumber
\end{align}
it can be seen that a solution to (\ref{CasOp}) is given by the expression%
\begin{equation}
L^{2}\equiv g_{AB}L^{A}L^{B}=-qL^{+}L^{-}+L^{3}L^{3}-q^{-1}L^{-}L^{+},
\label{L2}%
\end{equation}
which, in fact, is a q-analog for the Casimir operator of classical angular
momentum algebra. Equivalently, the Casimir operator for $U_{q}(su_{2})$ can
also be written in terms of the $M^{\prime}s$. In this manner, it becomes%
\begin{align}
M^{2}  &  \equiv g_{AB}g_{CD}M^{AC}M^{BD}\\
&  =q^{-1}M^{+3}M^{3-}-M^{+-}M^{+-}+q^{-3}M^{3-}M^{+3},\nonumber
\end{align}
which agrees with $L^{2}$ up to a constant factor.

Before closing this section, we wish to specify our Casimir operator for the
different representations addressed so far. Substituting for the symmetry
generators their representations we finally obtain

\begin{enumerate}
\item[a)] (operator representation)%
\begin{equation}
L^{2}=-(X\circ X)(\partial\circ\partial)+q^{2}(X\circ\partial)(X\circ
\partial)+q^{-2}X\circ\partial,
\end{equation}
with $U\circ V\equiv g_{AB}U^{A}V^{B}$,

\item[b)] (spinor representation)%
\begin{equation}
L^{2}=q^{-2}\lambda_{+}^{-2}{[}{[}3{]}{]}_{q^{2}}%
\mbox{1 \kern-.59em {\rm l}}_{2\times2},
\end{equation}

\item[c)] (vector representation)%
\begin{equation}
L^{2}=q^{-2}{[}{[}2{]}{]}_{q^{4}}\mbox{1 \kern-.59em {\rm l}}_{3\times3},
\end{equation}

\end{enumerate}

\noindent where the antisymmetric q-numbers are defined by
\begin{equation}
{[[}n{]]}_{q}\equiv\frac{1-q^{an}}{1-q^{a}},\quad n\in\mathbb{N},\quad
a\in\mathbb{C}.
\end{equation}

\section{Quantum Lie algebra of four-dimensional angular momentum}

As next example we would like to consider the q-deformed algebra of
four-dimensional angular momentum. This case can be treated in very much the
same way as the three-dimensional one. Thus, we restrict ourselves to stating
the results, only.

\subsection{Representation of four-dimensional angular momentum within
q-deformed differential calculus}

First of all, let us start with the defining representation. Within the
differential calculus the generators of q-deformed four-dimensional angular
momentum are represented by (see for example Ref. \cite{Oca96})
\begin{equation}
L^{ij}\equiv-q^{-2}\lambda_{+}\Lambda^{1/2}(P_{A})^{ij}{}_{kl}\,X^{k}%
\hat{\partial}^{l}=\lambda_{+}\Lambda^{1/2}(P_{A})^{ij}{}_{kl}\,\hat{\partial
}^{k}X^{l},
\end{equation}
or
\begin{equation}
L^{ij}\equiv-\lambda_{+}\Lambda^{-1/2}(P_{A})^{ij}{}_{kl}\,X^{k}\partial
^{l}=q^{-2}\lambda_{+}\Lambda^{-1/2}(P_{A})^{ij}{}_{kl}\,\partial^{k}X^{l},
\label{Def4dimAngMom}%
\end{equation}
where $P_{A}$ is a q-analog of an antisymmetrizer and $\Lambda$ denotes a
scaling operator subject to
\begin{equation}
\Lambda X^{i}=q^{2}X^{i}\Lambda,\quad\Lambda\partial^{i}=q^{-2}\partial
^{i}\Lambda.
\end{equation}
Eq. (\ref{Def4dimAngMom}) shows us that the $L^{ij}$ are components of an
antisymmetric tensor operator. More specifically, they satisfy
\begin{gather}
L^{ii}=0,\quad i=1,\ldots,4,\\
L^{j1}=-qL^{1j},\quad L^{4j}=-qL^{j4},\quad j=2,3,\nonumber\\
L^{41}=-L^{14},\quad L^{32}=-L^{23}+\lambda L^{14}.\nonumber
\end{gather}
Thus, we have in complete analogy to the classical case only six independent
generators, for which we can choose the set of $L^{ij}$ with $i<j.$ Taking the
explicit form of $P_{A}$ into consideration, we get for them more explicitly
\begin{align}
L^{1i}  &  =-\Lambda^{-1/2}(q^{-1}X^{1}\hat{\partial}^{i}-X^{i}\hat{\partial
}^{1}),\quad i=1,2,\\
L^{i4}  &  =-\Lambda^{-1/2}(q^{-1}X^{i}\hat{\partial}^{4}-X^{4}\hat{\partial
}^{i}),\nonumber\\
L^{14}  &  =2\lambda_{+}^{-1}\Lambda^{-1/2}(X^{1}\hat{\partial}^{4}-X^{4}%
\hat{\partial}^{1})\nonumber\\
&  -\;\lambda\lambda_{+}^{-1}\Lambda^{-1/2}(X^{2}\hat{\partial}^{3}+X^{3}%
\hat{\partial}^{2}),\nonumber\\
L^{23}  &  =(q^{2}+q^{-2})\lambda_{+}^{-1}\Lambda^{-1/2}X^{2}\hat{\partial
}^{3}\nonumber\\
&  -\;2\lambda_{+}^{-1}\Lambda^{-1/2}X^{3}\hat{\partial}^{2}+\lambda
\lambda_{+}^{-1}\Lambda^{-1/2}(X^{1}\hat{\partial}^{4}-X^{4}\hat{\partial}%
^{1}).\nonumber
\end{align}

Together with two grouplike braiding operators $K_{i},i=1,2$, the generators
$L^{kl},$ $k<l$, span the Hopf algebra $U_{q}(so_{4})$. Applying the
commutation relations for coordinates, partial derivatives and the two
braiding operators (see for example Appendix \ref{AppA} and Ref. \cite{BW01})
leaves us with the nontrivial relations%
\begin{gather}
L^{14}L^{1j}-L^{1j}L^{14}=-K_{i}L^{12},\quad(i,j)\in
\{(2,2),(1,3)\},\label{UqSO4Anf}\\
L^{l4}L^{14}-L^{14}L^{l4}=-q^{-2}K_{k}L^{l4},\quad(k,l)\in
\{(1,2),(2,3)\},\nonumber\\[0.16in]
L^{23}L^{12}-L^{12}L^{23}=-qK_{2}L^{12},\quad L^{23}L^{13}-L^{13}L^{23}%
=q^{-1}K_{1}L^{13},\\
L^{24}L^{23}-L^{23}L^{24}=q^{-3}K_{1}L^{24},\quad L^{34}L^{23}-L^{23}%
L^{34}=-q^{-1}K_{2}L^{34},\nonumber\\[0.16in]
K_{1}L^{13}=q^{-2}L^{13}K_{1},\quad K_{2}L^{12}=q^{-2}L^{12}K_{2}%
,\label{Uqsu4End}\\
K_{1}L^{24}=q^{2}L^{24}K_{1},\quad K_{2}L^{34}=q^{2}L^{34}K_{2}.\nonumber
\end{gather}

\subsection{Hopf structure of $U_{q}(so_{4})$ and corresponding q-commutators}

As in the three-dimensional case we need the Hopf structure to proceed
further. For the independent generators this reads as (see also Ref.
\cite{Oca96})%
\begin{align}
\Delta(L^{1j})  &  =L^{1j}\otimes K_{i}+1\otimes L^{12},\quad(i,j)\in
\{(2,2),(1,3)\},\\
\Delta(L^{l4})  &  =L^{l4}\otimes K_{k}+1\otimes L^{l4},\quad(k,l)\in
\{(1,2),(2,3)\},\nonumber\\
\Delta(L^{14})  &  =q\lambda_{+}^{-1}L^{14}\otimes K_{1}+q^{-1}\lambda
_{+}^{-1}L^{14}\otimes K_{2}\nonumber\\
&  +\;q\lambda_{+}^{-1}K_{1}^{-1}\otimes L^{14}+q^{-1}\lambda_{+}^{-1}%
K_{2}^{-1}\otimes L^{14}\nonumber\\
&  -\;\lambda_{+}^{-1}L^{23}\otimes K_{1}+\lambda_{+}^{-1}L^{23}\otimes
K_{2}\nonumber\\
&  -\;\lambda_{+}^{-1}K_{1}^{-1}\otimes L^{23}+\lambda_{+}^{-1}K_{2}%
^{-1}\otimes L^{23}\nonumber\\
&  -\;q\lambda\lambda_{+}^{-1}L^{24}K_{1}^{-1}\otimes L^{13}-q\lambda
\lambda_{+}^{-1}K_{1}^{-1}L^{13}\otimes L^{24}\nonumber\\
&  -\;q\lambda\lambda_{+}^{-1}L^{34}K_{2}^{-1}\otimes L^{12}-q\lambda
\lambda_{+}^{-1}K_{2}^{-1}L^{12}\otimes L^{34},\nonumber\\
\Delta(L^{23})  &  =-\lambda_{+}^{-1}K_{1}^{-1}\otimes L^{14}+\lambda_{+}%
^{-1}K_{2}^{-1}\otimes L^{14}\nonumber\\
&  -\;\lambda_{+}^{-1}L^{14}\otimes K_{1}+\lambda_{+}^{-1}L^{14}\otimes
K_{2}\nonumber\\
&  +\;q^{-1}\lambda_{+}^{-1}L^{23}\otimes K_{1}+q\lambda_{+}^{-1}L^{23}\otimes
K_{2}\nonumber\\
&  +\;q^{-1}\lambda_{+}^{-1}K_{1}^{-1}\otimes L^{23}+q\lambda_{+}^{-1}%
K_{2}^{-1}\otimes L^{23}\nonumber\\
&  +\;\lambda\lambda_{+}^{-1}L^{24}K_{1}^{-1}\otimes L^{13}+\lambda\lambda
_{+}^{-1}K_{1}^{-1}L^{13}\otimes L^{24}\nonumber\\
&  -\;q^{2}\lambda\lambda_{+}^{-1}L^{34}K_{2}^{-1}\otimes L^{12}-q^{2}%
\lambda\lambda_{+}^{-1}K_{2}^{-1}L^{12}\otimes L^{34},\nonumber\\[0.16in]
S(L^{1j})  &  =q^{-2}S^{-1}(L^{1j})=-L^{1j}K_{i}^{-1},\quad(i,j)\in
\{(2,2),(1,3)\},\\
S(L^{l4})  &  =q^{2}S^{-1}(L^{l4})=-L^{l4}K_{k}^{-1},\quad(k,l)\in
\{(1,2),(2,3)\},\nonumber\\
S(L^{14})  &  =S^{-1}(L^{14})=L^{14}-q^{-1}\lambda^{-1}((K_{1}+K_{2}%
)-(K_{1}^{-1}+K_{2}^{-1})),\nonumber\\
S(L^{23})  &  =S^{-1}(L^{23})=L^{23}+q^{-1}\lambda^{-1}(q^{-1}(K_{1}%
-K_{1}^{-1})-q(K_{2}-K_{2}^{-1})),\nonumber\\[0.16in]
\varepsilon(L^{mn})  &  =0.
\end{align}

Again, this Hopf structure enables us to introduce q-commutators in complete
analogy to the three-dimensional case. Explicitly written out we obtain for
the q-commutator with an element living in a representation space of
$U_{q}(so_{4})$%
\begin{align}
{[}L^{1j},V{]}_{q}  &  =(L^{1j}V-VL^{1j})K_{i}^{-1},\quad(i,j)\in
\{(2,2),(1,3)\},\label{commeu4l}\\
{[}L^{l4},V{]}_{q}  &  =(L^{l4}V-VL^{l4})K_{k}^{-1},\quad(k,l)\in
\{(1,2),(2,3)\},\nonumber\\
{[}L^{14},V{]}_{q}  &  =-q^{-1}\lambda^{-1}(K_{1}^{-1}VK_{1}+K_{2}^{-1}%
VK_{2})\nonumber\\
&  +\;q^{-1}\lambda^{-1}(K_{1}^{-1}VK_{1}^{-1}+K_{2}^{-1}VK_{2}^{-1}%
)\nonumber\\
&  -\;\lambda_{+}^{-1}(K_{2}^{-1}VL^{23}+L^{23}VK_{2}^{-1}-K_{1}^{-1}%
VL^{23}+L^{23}VK_{1}^{-1})\nonumber\\
&  +\;q^{-1}\lambda_{+}^{-1}(K_{2}^{-1}VL^{14}+L^{14}VK_{2}^{-1})\nonumber\\
&  +\;q\lambda_{+}^{-1}(K_{1}^{-1}VL^{14}+L^{14}VK_{1}^{-1})\nonumber\\
&  +\;q\lambda\lambda_{+}^{-1}(K_{1}^{-1}L^{13}VL^{24}K_{1}^{-1}+K_{2}%
^{-1}L^{12}VL^{34}K_{2}^{-1})\nonumber\\
&  +\;q\lambda\lambda_{+}^{-1}(L^{24}K_{1}^{-1}VL^{13}K_{1}^{-1}+L^{34}%
K_{2}^{-1}VL^{12}K_{2}^{-1}),\nonumber\\
{[}L^{23},V{]}_{q}  &  =q\lambda_{+}^{-1}(K_{2}^{-1}VL^{23}+L^{23}VK_{2}%
^{-1})\nonumber\\
&  +\;q^{-1}\lambda_{+}^{-1}(K_{1}^{-1}VL^{23}+L^{23}VK_{1}^{-1})\nonumber\\
&  +\;\lambda_{+}^{-1}(L^{14}VK_{2}^{-1}+K_{2}^{-1}VL^{14}-L^{14}VK_{1}%
^{-1}-K_{1}^{-1}VL^{14})\nonumber\\
&  +\;\lambda^{-1}(K_{2}^{-1}VK_{2}^{-1}-K_{2}^{-1}VK_{2})\nonumber\\
&  +\;q^{-2}\lambda^{-1}(K_{1}^{-1}VK_{1}-K_{1}^{-1}VK_{1}^{-1})\nonumber\\
&  +\;q^2\lambda\lambda_{+}^{-1}(K_{2}^{-1}L^{12}%
VL^{34}K_{2}^{-1}+L^{34}K_{2}^{-1}VL^{12}K_{2}%
^{-1})\nonumber\\
&  -\;\lambda\lambda_{+}^{-1}(K_{1}^{-1}L^{13}VL^{24}K_{1}^{-1}-L^{24}%
K_{1}^{-1}VL^{13}K_{1}^{-1}),\nonumber
\end{align}
and likewise for their right versions,
\begin{align}
{[}V,L^{1j}{]}_{q}  &  =K_{i}^{-1}(VL^{1j}-L^{1j}V),\quad(i,j)\in
\{(2,2),(1,3)\},\label{commeu4r}\\
{[}V,L^{l4}{]}_{q}  &  =K_{k}^{-1}(VL^{l4}-L^{l4}V),\quad(k,l)\in
\{(1,2),(2,3)\},\nonumber\\
{[}V,L^{14}{]}_{q}  &  =q\lambda_{+}^{-1}(L^{14}VK_{1}^{-1}+K_{1}^{-1}%
VL^{14})\nonumber\\
&  +\;q^{-1}\lambda_{+}^{-1}(L^{14}VK_{2}^{-1}+K_{2}^{-1}VL^{14})\nonumber\\
&  +\;\lambda_{+}^{-1}(L^{23}VK_{2}^{-1}+K_{2}^{-1}VL^{23}-L^{23}VK_{1}%
^{-1}-K_{1}^{-1}VL^{23})\nonumber\\
&  +\;q^{-1}\lambda^{-1}(K_{1}^{-1}VK_{1}^{-1}-K_{1}VK_{1}^{-1})\nonumber\\
&  +\;q^{-1}\lambda^{-1}(K_{2}^{-1}VK_{2}^{-1}-K_{2}VK_{2}^{-1})\nonumber\\
&  +\;q\lambda\lambda_{+}^{-1}(K_{1}^{-1}L^{13}VL^{24}K_{1}^{-1}+K_{1}%
^{-1}L^{24}VK_{1}^{-1}L^{13})\nonumber\\
&  +\;q\lambda\lambda_{+}^{-1}(K_{2}^{-1}L^{34}VK_{2}^{-1}L^{12}+K_{2}%
^{-1}L^{12}VL^{34}K_{2}^{-1}),\nonumber\\
{[}V,L^{23}{]}_{q}  &  =q\lambda_{+}^{-1}(L^{23}VK_{2}^{-1}+K_{2}^{-1}%
VL^{23})\nonumber\\
&  +\;q^{-1}\lambda_{+}^{-1}(L^{23}VK_{1}^{-1}+K_{1}^{-1}VL^{23})\nonumber\\
&  +\;\lambda_{+}^{-1}(L^{14}VK_{2}^{-1}+K_{2}^{-1}VL^{14}-L^{14}VK_{1}%
^{-1}-K_{1}^{-1}VL^{14})\nonumber\\
&  +\;\lambda^{-1}(K_{2}^{-1}VK_{2}^{-1}-K_{2}VK_{2}^{-1})\nonumber\\
&  +\;q^{-2}\lambda^{-1}(K_{1}VK_{1}^{-1}-K_{1}^{-1}VK_{1}^{-1})\nonumber\\
&  +\;\lambda\lambda_{+}^{-1}(q^{2}K_{2}^{-1}L^{34}VK_{2}^{-1}L^{12}%
+q^2K_{2}^{-1}L^{12}VL^{34}K_{2}^{-1})\nonumber\\
&  -\;\lambda\lambda_{+}^{-1}(K_{1}^{-1}L^{13}VL^{24}K_{1}^{-1}-K_{1}%
^{-1}L^{24}VK_{1}^{-1}L^{13}).\nonumber
\end{align}

\subsection{Matrix representations of $U_{q}(so_{4})$ and commutation
relations with tensor operators}

Let us now go on to the spinor and vector representations of the $L^{ij}$. As
in the classical case, we can distinguish two types of spinor representations,
i.e. $(1/2,0)$ and $(0,1/2)$ (see for example Ref. \cite{Oca96}). Explicitly,
we have
\begin{align}
L^{ij}\rhd\theta^{\,\alpha}  &  =(\sigma^{ij})^{\alpha}{}_{\beta}%
\,\theta^{\beta}, & \quad L^{ij}\rhd\tilde{\theta}^{\,\alpha}  &
=(\tilde{\sigma}^{ij})^{\alpha}{}_{\beta}\,\tilde{\theta}^{\beta}\\
\theta_{\alpha}\lhd L^{ij}  &  =\theta_{\beta}(\sigma^{ij})^{\beta}{}_{\alpha
}, & \quad\tilde{\theta}_{\alpha}\lhd L^{ij}  &  =\tilde{\theta}_{\beta
}(\tilde{\sigma}^{ij})^{\beta}{}_{\alpha},\nonumber
\end{align}
with
\begin{gather}
(\sigma^{13})^{\alpha}{}_{\beta}=\left(
\begin{array}
[c]{cc}%
0 & -q^{-2}\\
0 & 0
\end{array}
\right)  ,\qquad(\sigma^{24})^{\alpha}{}_{\beta}=\left(
\begin{array}
[c]{cc}%
0 & 0\\
-q^{-1} & 0
\end{array}
\right)  ,\\
(\sigma^{14})^{\alpha}{}_{\beta}=-q(\sigma^{23})^{\alpha}{}_{\beta}%
=q^{-1}\lambda_{+}^{-1}\left(
\begin{array}
[c]{cc}%
-q & 0\\
0 & q^{-1}%
\end{array}
\right)  ,\nonumber\\
(\sigma^{34})^{\alpha}{}_{\beta}=(\sigma^{12})^{\alpha}{}_{\beta}=0.\nonumber
\end{gather}
The matrices with tilde one gets most easily from the identities
\begin{align}
(\tilde{\sigma}^{13})^{\alpha}{}_{\beta}  &  =(\sigma^{12})^{\alpha}{}_{\beta
}, & (\tilde{\sigma}^{12})^{\alpha}{}_{\beta}  &  =(\sigma^{13})^{\alpha}%
{}_{\beta},\\
(\tilde{\sigma}^{14})^{\alpha}{}_{\beta}  &  =(\sigma^{14})^{\alpha}{}_{\beta
}, & (\tilde{\sigma}^{23})^{\alpha}{}_{\beta}  &  =-q^{-2}(\sigma
^{23})^{\alpha}{}_{\beta},\nonumber\\
(\tilde{\sigma}^{34})^{\alpha}{}_{\beta}  &  =(\sigma^{24})^{\alpha}{}_{\beta
}, & (\tilde{\sigma}^{24})^{\alpha}{}_{\beta}  &  =(\sigma^{34})^{\alpha}%
{}_{\beta}.\nonumber
\end{align}
Notice that the tilde on the spinor components shall remind us of the fact
that they transformed differently. That these matrices determine a
representation of $U_{q}(so_{4})$ can again be proven by inserting them
together with%
\begin{gather}
(K_{1})^{\alpha}{}_{\beta}=\left(
\begin{array}
[c]{cc}%
q & 0\\
0 & q^{-1}%
\end{array}
\right)  ,\qquad(\tilde{K}_{2})^{\alpha}{}_{\beta}=\left(
\begin{array}
[c]{cc}%
q^{-1} & 0\\
0 & q
\end{array}
\right)  ,\\
(\tilde{K}_{1})^{\alpha}{}_{\beta}=(K_{2})^{\alpha}{}_{\beta}%
=\mbox{1 \kern-.59em {\rm l}},\nonumber
\end{gather}
into relations (\ref{UqSO4Anf}) - (\ref{Uqsu4End}).

From the spinor representation of $U_{q}(so_{4})$ we can - as usual -\ compute
commutation relations between the $L^{ij}$ and the components of a spinor
operator. For this to achieve we apply the identities
\begin{equation}
\left[  L^{ij},\theta^{\,\alpha}\right]  _{q}=(\sigma^{ij})^{\alpha}{}_{\beta
}\,\theta^{\beta},\quad\left[  \theta^{\,\alpha},L^{ij}\right]  _{q}%
=\theta_{\beta}\,(\sigma^{ij})^{\beta}{}_{\alpha}.
\end{equation}
If we write out the q-commutators and substitute the representation matrices
without tilde for the $L^{ij}$ we get the commutation relations
\begin{align}
L^{13}\theta^{2}  &  =\theta^{2}L^{13}-q^{-2}\theta^{1}K_{1},\\
L^{24}\theta^{1}  &  =\theta^{1}L^{24}-q^{-1}\theta^{2}K_{1},\nonumber\\
L^{14}\theta^{1}  &  =(q^{2}+q^{-1})\lambda_{+}^{-1}\theta^{1}L^{14}%
-(q-1)\lambda_{+}^{-1}\theta^{1}L^{23}\nonumber\\
&  +\;q\lambda\lambda_{+}^{-1}\theta^{2}L^{13}-\lambda_{+}^{-1}\theta^{1}%
K_{1},\nonumber\\
L^{14}\theta^{2}  &  =(q^{-1}+1)\lambda_{+}^{-1}\theta^{2}L^{14}%
+(q^{-1}-1)\lambda_{+}^{-1}\theta^{2}L^{23}\nonumber\\
&  +\;\lambda\lambda_{+}^{-1}\theta^{1}L^{24}+q^{-2}\lambda_{+}^{-1}\theta
^{2}K_{1},\nonumber\\
L^{23}\theta^{1}  &  =(q+1)\lambda_{+}^{-1}\theta^{1}L^{23}-(q-1)\lambda
_{+}^{-1}\theta^{1}L^{14}\nonumber\\
&  -\;\lambda\lambda_{+}^{-1}\theta^{2}L^{13}+q^{-1}\lambda_{+}^{-1}\theta
^{1}K_{1},\nonumber\\
L^{23}\theta^{2}  &  =(q^{-2}+q)\lambda_{+}^{-1}\theta^{2}L^{23}%
+(1-q^{-1})\lambda_{+}^{-1}\theta^{2}L^{14}\nonumber\\
&  -\;q^{-1}\lambda\lambda_{+}^{-1}\theta^{1}L^{24}-q^{-3}\lambda_{+}%
^{-1}\theta^{2}K_{1},\nonumber
\end{align}
and%
\begin{align}
\theta_{1}L^{13}  &  =L^{13}\theta_{1}-q^{-2}K_{1}\theta_{2},\\
\theta_{2}L^{24}  &  =L^{24}\theta_{2}-q^{-1}K_{1}\theta_{1},\nonumber\\
\theta_{1}L^{14}  &  =(q^{2}+q^{-1})\lambda_{+}^{-1}L^{14}\theta
_{1}-(q-1)\lambda_{+}^{-1}L^{23}\theta_{1}\nonumber\\
&  +\;\lambda\lambda_{+}^{-1}L^{24}\theta_{2}-\lambda_{+}^{-1}K_{1}\theta
_{1},\nonumber\\
\theta_{2}L^{14}  &  =(q^{-1}+1)\lambda_{+}^{-1}L^{14}\theta_{2}%
+(1-q^{-1})\lambda_{+}^{-1}L^{23}\theta_{2}\nonumber\\
&  +\;q\lambda\lambda_{+}^{-1}L^{13}\theta_{1}+q^{-2}\lambda_{+}^{-1}%
K_{1}\theta_{2},\nonumber\\
\theta_{1}L^{23}  &  =(q+1)\lambda_{+}^{-1}L^{23}\theta_{1}-(q-1)\lambda
_{+}^{-1}L^{14}\theta_{1}\nonumber\\
&  -\;q^{-1}\lambda\lambda_{+}^{-1}L^{24}\theta_{2}+q^{-1}\lambda_{+}%
^{-1}K_{1}\theta_{1},\nonumber\\
\theta_{2}L^{23}  &  =(q+q^{-2})\lambda_{+}^{-1}L^{23}\theta_{2}%
+(1-q^{-1})\lambda_{+}^{-1}L^{14}\theta_{2}\nonumber\\
&  -\;\lambda\lambda_{+}^{-1}L^{13}\theta_{1}-q^{-3}\lambda_{+}^{-1}%
K_{1}\theta_{2}.\nonumber
\end{align}
The remaining relations are trivial, i.e. these relations take the form
$L^{ij}\theta^{k}=\theta^{k}L^{ij}.$ Repeating the same steps as before for
spinors (and the representation matrices) with tilde we arrive at the
following nontrivial relations:
\begin{align}
L^{12}\tilde{\theta}^{2}  &  =\tilde{\theta}^{2}L^{12}-q^{-2}\tilde{\theta
}^{1}K_{2},\\
L^{34}\tilde{\theta}^{1}  &  =\tilde{\theta}^{1}L^{34}-q^{-1}\tilde{\theta
}^{2}K_{2},\nonumber\\
L^{14}\tilde{\theta}^{1}  &  =(q+1)\lambda_{+}^{-1}\tilde{\theta}^{1}%
L^{14}+(q-1)\lambda_{+}^{-1}\tilde{\theta}^{1}L^{23}\nonumber\\
&  +\;q\lambda\lambda_{+}^{-1}\tilde{\theta}^{2}L^{12}-\lambda_{+}^{-1}%
\tilde{\theta}^{1}K_{2},\nonumber\\
L^{14}\tilde{\theta}^{2}  &  =(q^{-2}+q)\lambda_{+}^{-1}\tilde{\theta}%
^{2}L^{12}+(q^{-1}-1)\lambda_{+}^{-1}\tilde{\theta}^{2}L^{23}\nonumber\\
&  +\;\lambda\lambda_{+}^{-1}\tilde{\theta}^{1}L^{34}+q^{-2}\lambda_{+}%
^{-1}\tilde{\theta}^{2}K_{2},\nonumber\\
L^{23}\tilde{\theta}^{1}  &  =(q^{-1}+q^{2})\lambda_{+}^{-1}\tilde{\theta}%
^{1}L^{23}+(q-1)\lambda_{+}^{-1}\tilde{\theta}^{1}L^{14}\nonumber\\
&  +\;q^{2}\lambda\lambda_{+}^{-1}\tilde{\theta}^{2}L^{12}-q\lambda_{+}%
^{-1}K_{2},\nonumber\\
L^{23}\tilde{\theta}^{2}  &  =(q^{-1}+1)\lambda_{+}^{-1}\tilde{\theta}%
^{2}L^{23}+(q^{-1}-1)\lambda_{+}^{-1}\tilde{\theta}^{2}L^{14}\nonumber\\
&  +\;q\lambda\lambda_{+}^{-1}\tilde{\theta}^{1}L^{34}+q^{-1}\lambda_{+}%
^{-1}\tilde{\theta}^{2}K_{2},\nonumber
\end{align}
and
\begin{align}
\tilde{\theta}_{1}L^{12}  &  =L^{12}\tilde{\theta}_{1}-q^{-2}K_{2}%
\tilde{\theta}_{2},\\
\tilde{\theta}_{2}L^{34}  &  =L^{34}\tilde{\theta}_{2}-q^{-1}K_{2}%
\tilde{\theta}_{1},\nonumber\\
\tilde{\theta}_{1}L^{14}  &  =(q+1)\lambda_{+}^{-1}L^{14}\tilde{\theta}%
_{1}+(q-1)\lambda_{+}^{-1}L^{23}\tilde{\theta}_{1}\nonumber\\
&  +\;\lambda\lambda_{+}^{-1}L^{34}\tilde{\theta}_{2}-\lambda_{+}^{-1}%
K_{2}\tilde{\theta}_{1},\nonumber\\
\tilde{\theta}_{2}L^{14}  &  =(q+q^{-2})\lambda_{+}^{-1}L^{14}\tilde{\theta
}_{2}+(q^{-1}-1)\lambda_{+}^{-1}L^{23}\tilde{\theta}_{2}\nonumber\\
&  +\;q\lambda\lambda_{+}^{-1}L^{12}\tilde{\theta}_{1}+q^{-2}\lambda_{+}%
^{-1}K_{2}\tilde{\theta}_{2},\nonumber\\
\tilde{\theta}_{1}L^{23}  &  =(q^{2}+q^{-1})\lambda_{+}^{-1}L^{23}%
\tilde{\theta}_{1}+(q-1)\lambda_{+}^{-1}L^{14}\tilde{\theta}_{1}\nonumber\\
&  +\;q\lambda\lambda_{+}^{-1}L^{34}\tilde{\theta}_{2}-q\lambda_{+}^{-1}%
K_{2}\tilde{\theta}_{1},\nonumber\\
\tilde{\theta}_{2}L^{23}  &  =(1+q^{-1})\lambda_{+}^{-1}L^{23}\tilde{\theta
}_{2}+(q^{-1}-1)\lambda_{+}^{-1}L^{14}\tilde{\theta}_{2}\nonumber\\
&  +\;q^{2}\lambda\lambda_{+}^{-1}L^{12}\tilde{\theta}_{1}+q^{-1}\lambda
_{+}^{-1}K_{2}\tilde{\theta}_{2}.\nonumber
\end{align}

These considerations carry over to the\ vector representations of the
independent generators of $U_{q}(so_{4})$. Its right and left versions are
given by
\begin{equation}
L^{ij}\rhd X^{k}=(\tau^{ij})^{k}{}_{m}\,X^{m},\quad X_{k}\lhd L^{ij}%
=X_{m}((\tau^{ij})^{m}{}_{k},
\end{equation}
with
\begin{gather}
(\tau^{14})^{k}{}_{m}=q^{-1}\lambda_{+}^{-1}\left(
\begin{array}
[c]{cccc}%
-2q & 0 & 0 & 0\\
0 & -\lambda & 0 & 0\\
0 & 0 & -\lambda & 0\\
0 & 0 & 0 & 2q^{-1}%
\end{array}
\right)  ,\\
(\tau^{12})^{k}{}_{m}=q^{-2}\left(
\begin{array}
[c]{cccc}%
0 & 0 & -1 & 0\\
0 & 0 & 0 & 1\\
0 & 0 & 0 & 0\\
0 & 0 & 0 & 0
\end{array}
\right)  ,\quad(\tau^{13})^{k}{}_{m}=q^{-2}\left(
\begin{array}
[c]{cccc}%
0 & -1 & 0 & 0\\
0 & 0 & 0 & 0\\
0 & 0 & 0 & 1\\
0 & 0 & 0 & 0
\end{array}
\right)  ,\nonumber\\
(\tau^{24})^{k}{}_{m}=q^{-1}\left(
\begin{array}
[c]{cccc}%
0 & 0 & 0 & 0\\
-1 & 0 & 0 & 0\\
0 & 0 & 0 & 0\\
0 & 0 & 1 & 0
\end{array}
\right)  ,\quad(\tau^{34})^{k}{}_{m}=q^{-1}\left(
\begin{array}
[c]{cccc}%
0 & 0 & 0 & 0\\
0 & 0 & 0 & 0\\
-1 & 0 & 0 & 0\\
0 & 1 & 0 & 0
\end{array}
\right)  ,\nonumber\\
(\tau^{23})^{k}{}_{m}=q^{-1}\lambda_{+}^{-1}\left(
\begin{array}
[c]{cccc}%
-q\lambda & 0 & 0 & 0\\
0 & -(q^{2}+q^{-2}) & 0 & 0\\
0 & 0 & 2 & 0\\
0 & 0 & 0 & q^{-1}\lambda
\end{array}
\right)  .\nonumber
\end{gather}
If we want to check that the above matrices give a representation of
$U_{q}(so_{4})$ we additionally need the matrices corresponding to the
braiding operators, i.e.
\begin{equation}
(K_{1})^{k}{}_{m}=\left(
\begin{array}
[c]{cccc}%
q^{-1} & 0 & 0 & 0\\
0 & q & 0 & 0\\
0 & 0 & q^{-1} & 0\\
0 & 0 & 0 & q
\end{array}
\right)  ,\quad(K_{2})^{k}{}_{m}=\left(
\begin{array}
[c]{cccc}%
q^{-1} & 0 & 0 & 0\\
0 & q^{-1} & 0 & 0\\
0 & 0 & q & 0\\
0 & 0 & 0 & q
\end{array}
\right)  .
\end{equation}

We are again in a position to write down commutation relations between the
independent $L^{ij}$ and the components of a vector operator. Starting from%
\begin{equation}
{[}L^{ij},X^{k}{]}_{q}=(\tau^{ij})^{k}{}_{l}\,X^{l}, \label{73}%
\end{equation}
and proceeding in very much the same way as in the spinor case we obtain the relations%

\begin{align}
L^{1j}X^{1}  &  =X^{1}L^{1j},\\
L^{1j}X^{j}  &  =X^{j}L^{1j},\nonumber\\
L^{1k}X^{l}  &  =X^{l}L^{1k}-q^{-2}X^{1}K_{i},\nonumber\\
L^{1m}X^{4}  &  =X^{4}L^{1m}+q^{-2}X^{m}K_{n},\nonumber\\[0.16in]
L^{m^{\prime}4}X^{1}  &  =X^{1}L^{m^{\prime}4}-q^{-1}X^{m^{\prime}}K_{n},\\
L^{j4}X^{j}  &  =X^{j}L^{j4},\nonumber\\
L^{l4}X^{k}  &  =X^{k}L^{l4}+q^{-1}X^{4}K_{i},\nonumber\\
L^{j4}X^{4}  &  =X^{4}L^{j4},\nonumber
\end{align}
{and}
\begin{align}
L^{14}X^{1}  &  =qX^{1}L^{14}-\lambda_{+}^{-1}X^{1}(K_{1}+K_{2})\\
&  +\;q\lambda\lambda_{+}^{-1}X^{2}%
L^{13}+X^{3}L^{12}),\nonumber\\
L^{14}X^{2}  &  =2\lambda_{+}^{-1}X^{2}L^{14}+\lambda_{+}^{-1}X^{2}(q^{2}%
K_{1}-K_{2})+\lambda\lambda_{+}^{-1}X^{2}L^{23}\nonumber\\
&  +\;\lambda\lambda_{+}^{-1}(X^{1}L^{24}-qX^{4}L^{12}),\nonumber\\
L^{14}X^{3}  &  =q^{-2}(q^{4}+1)\lambda_{+}^{-1}X^{3}L^{14}+\lambda_{+}%
^{-1}X^{3}(q^{-2}K_{2}-K_{1}%
)\nonumber\\
&  -\;\lambda\lambda_{+}^{-1}X^{3}L^{23}+\lambda\lambda_{+}^{-1}(X^{1}%
L^{34}-qX^{4}L^{13}),\nonumber\\
L^{14}X^{4}  &  =q^{-1}X^{4}L^{14}+q^{-2}\lambda_{+}^{-1}X^{4}(K_{1}%
+K_{2})\nonumber\\
&  -\;\lambda\lambda_{+}^{-1}(X^{2}L^{34}+X^{3}L^{24}),\nonumber\\[0.16in]
L^{23}X^{1}  &  =qX^{1}L^{23}+\lambda_{+}^{-1}X^{1}(q^{-1}K_{1}-qK_{2})\\
&  -\;\lambda\lambda_{+}^{-1}(X^{2}L^{13}-q^{2}X^{3}L^{12}),\nonumber\\
L^{23}X^{2}  &  =q^{-2}(q^{4}+1)X^{2}L^{23}-\lambda_{+}^{-1}X^{2}(q^{-3}%
K_{1}+qK_{2})\nonumber\\
&  -\;q^{2}\lambda_+^{-1} X^{4}L^{12}-\lambda\lambda_{+}^{-1}(q^{-1}X^{1}L^{24}%
+X^{2}L^{14}),\nonumber\\
L^{23}X^{3}  &  =2\lambda_{+}^{-1}X^{3}L^{23}+q^{-1}\lambda_{+}^{-1}%
X^{3}(K_{1}+K_{2})\nonumber\\
&  -\;\lambda\lambda_{+}^{-1}X^{3}L^{14}+\lambda\lambda_{+}^{-1}(qX^{1}%
L^{34}+X^{4}L^{13}),\nonumber\\
L^{23}X^{4}  &  =q^{-1}X^{4}L^{23}-q^{-1}\lambda_{+}^{-1}X^{4}(q^{-2}%
K_{1}-K_{2})\nonumber\\
&  -\;\lambda\lambda_{+}^{-1}(qX^{2}L^{34}-q^{-1}X^{3}L^{24}),\nonumber
\end{align}
where%
\begin{gather}
j=2,3,\quad(k,l,i)\in\{(2,3,2),(3,2,1)\},\\
(m,n)\in\{(2,2),(3,1)\},\quad m^{\prime}\equiv5-m^{\prime}.\nonumber
\end{gather}
The right version of (\ref{73}), i.e.
\begin{equation}
{[}X_{k},L^{ij}{]}_{q}=X_{l}\,(\tau^{ij})^{l}{}_{k}\,,
\end{equation}
gives us likewise%

\begin{align}
X_{1}L^{1m}  &  =L^{1m}X_{1}-q^{-2}K_{n}X_{3},\\
X_{l^{\prime}}L^{1k}  &  =L^{1k}X_{l^{\prime}}+q^{-2}K_{i}X_{4},\nonumber\\
X_{j^{\prime}}L^{1j}  &  =L^{1j}X_{j^{\prime}},\nonumber\\
X_{4}L^{1j}  &  =L^{1j}X_{4},\nonumber\\[0.16in]
X_{1}L^{j4}  &  =L^{j4}X_{1},\\
X_{k^{\prime}}L^{l4}  &  =L^{l4}X_{2}-q^{-1}K_{i}X_{1},\nonumber\\
X_{j^{\prime}}L^{j4}  &  =L^{j4}X_{j^{\prime}},\nonumber\\
X_{4}L^{m^{\prime}4}  &  =L^{m^{\prime}4}X_{4}+q^{-1}K_{n}X_{m},\nonumber
\end{align}
{and}
\begin{align}
X_{1}L^{14}  &  =qL^{14}X_{1}-\lambda_{+}^{-1}(K_{1}+K_{2})X_{1}\\
&  +\;\lambda\lambda_{+}^{-1}(L^{24}X_{2}+L^{34}X_{3}),\nonumber\\
X_{2}L^{14}  &  =2\lambda_{+}^{-1}L^{14}X_2+\lambda
_{+}^{-1}(q^{-2}K_{1}-K_{2})X_{2}\nonumber\\
&  +\;\lambda\lambda_{+}^{-1}L^{23}X_{2}-\lambda\lambda_{+}^{-1}(L^{34}%
X_{4}+qL^{13}X_{1}),\nonumber\\
X_{3}L^{14}  &  =q^{-2}(q^{4}+1)\lambda_{+}^{-1}L^{14}X_{3}+\lambda_{+}%
^{-1}(q^{-2}K_{2}-K_{1})X_{3}\nonumber\\
&  -\;\lambda\lambda_{+}^{-1}L^{23}X_{3}+\lambda\lambda_{+}^{-1}(qL^{12}%
X_{1}-L^{24}X_{4}),\nonumber\\
X_{4}L^{14}  &  =q^{-1}L^{14}X_{4}+q^{-2}\lambda_{+}^{-1}(K_{1}+K_{2}%
)X_{4}\nonumber\\
&  -\;q\lambda\lambda_{+}^{-1}(L^{12}X_{2}+L^{13}X_{3}),\nonumber\\[0.16in]
X_{1}L^{23}  &  =qL^{23}X_{1}+q^{-1}\lambda_{+}^{-1}(K_{1}-q^{2}K_{2})X_{1}\\
&  +\;\lambda\lambda_{+}^{-1}(qL^{34}X_{3}-q^{-1}L^{24}X_{2}),\nonumber\\
X_{2}L^{23}  &  =q^{-2}(q^{4}+1)L^{23}X_{2}-\lambda_{+}^{-1}(q^{-3}%
K_{1}+qK_{2})X_{2}\nonumber\\
&  +\;\lambda\lambda_{+}^{-1}L^{14}X_{2}-\lambda\lambda_{+}^{-1}(L^{13}%
X_{1}+qL^{34}X_{4}),\nonumber\\
X_{3}L^{23}  &  =2\lambda_{+}^{-1}L^{23}X_3+q^{-1}%
\lambda_{+}^{-1}(K_{1}+K_{2})X_{3}\nonumber\\
&  -\;\lambda\lambda_{+}^{-1}L^{14}X_{3}+\lambda\lambda_{+}^{-1}(q^{2}%
L^{12}X_{1}+q^{-1}L^{24}X_{4}),\nonumber\\
X_{4}L^{23}  &  =q^{-1}L^{23}X_4+q^{-1}\lambda_{+}%
^{-1}(K_{2}-q^{-2}K_{1})X_{4}\nonumber\\
&  +\;\lambda\lambda_{+}^{-1}(L^{13}X_{3}-q^{2}L^{12}X_{2}).\nonumber
\end{align}

\subsection{Quantum Lie algebra of $U_{q}(so_{4})$ and its Casimir operators}

The $L^{ij}$ can act on themselves through the adjoint representation, which
can be obtained via the following simple reasoning. We calculate the action of
the $L^{ij}$ on a product of antisymmetrised vector coordinates (for their
definition see Ref. \cite{Mik04}), i.e.
\begin{equation}
L^{ij}\triangleright\xi^{k}\xi^{l}=\left(  (L^{ij})_{(1)}\triangleright\xi
^{k}\right)  \left(  (L^{ij})_{(2)}\triangleright\xi^{l}\right)
=\sum\nolimits_{mn}(a^{ij,kl})_{mn}\,\xi^{m}\xi^{n}. \label{ActLijLkl}%
\end{equation}
As already mentioned, the $L^{ij}$ are components of an antisymmetric tensor
operator. Hence, they have to act on themselves in the same way as they act on
a product of antisymmetrised coordinates. In this sense we read off from
(\ref{ActLijLkl}) for their quantum Lie algebra
\begin{equation}
{[}L^{ij},L^{kl}{]}_{q}=L^{ij}\triangleright L^{kl}=\sum\nolimits_{mn}%
(a^{ij,kl})_{mn}\,L^{mn}.
\end{equation}
Explicitly, the non vanishing q-commutators for the independent $L^{ij}$
become
\begin{align}
{[}L^{12},L^{23}{]}_{q}  &  =-q^{-2}{[}L^{23},L^{12}{]}_{q}=q^{-1}L^{12},\\
{[}L^{12},L^{34}{]}_{q}  &  =-{[}L^{34},L^{12}{]}_{q}=-q^{-2}L^{23}%
-q^{-3}L^{14},\nonumber\\
{[}L^{13},L^{23}{]}_{q}  &  =-q^{-2}{[}L^{23},L^{13}{]}_{q}=-q^{-3}%
L^{13},\nonumber\\
{[}L^{13},L^{24}{]}_{q}  &  =-{[}L^{24},L^{13}{]}_{q}=q^{-2}L^{23}%
-q^{-1}L^{14},\nonumber\\[0.16in]
{[}L^{14},L^{1i}{]}_{q}  &  =-q^{2}{[}L^{1i},L^{14}{]}_{q}=-L^{1i},\quad
i=2,3,\\
{[}L^{14},L^{14}{]}_{q}  &  =-q^{-1}\lambda L^{14},\nonumber\\
{[}L^{14},L^{23}{]}_{q}  &  ={[}L^{23},L^{14}{]}_{q}=-q^{-1}\lambda
L^{23},\nonumber\\
{[}L^{14},L^{i4}{]}_{q}  &  =-q^{-2}{[}L^{i4},L^{14}{]}_{q}=q^{-2}L^{i4},\quad
i=2,3,\nonumber\\[0.16in]
{[}L^{23},L^{23}{]}_{q}  &  =-q^{-1}\lambda L^{14}-q^{-1}\lambda^{2}L^{23},\\
{[}L^{23},L^{24}{]}_{q}  &  =-q^{-2}{[}L^{24},L^{23}{]}_{q}=-q^{-3}%
L^{24},\nonumber\\
{[}L^{23},L^{34}{]}_{q}  &  =-q^{-2}{[}L^{34},L^{23}{]}_{q}=q^{-1}%
L^{34}.\nonumber
\end{align}
Let us remark that these identities can be checked by means of the spinor and
vector representations of the $L^{ij}$. Towards this end, one has to write out
the q-commutators using (\ref{commeu4l}) or (\ref{commeu4r}). Then we replace
the generators by their representation matrices and apply usual matrix
multiplication. Proceeding in this manner will again show us the validity of
the above identities.

At last, let us consider the Casimir operators, which can be introduced as in
the three-dimensional case. The first one is given by the expression
\begin{align}
C_{1}  &  =g_{ik}g_{jm}L^{ij}L^{km}=2L^{23}L^{23}+\lambda_{+}(L^{12}%
L^{34}+L^{13}L^{24})\\
&  =+q^{2}\lambda_{+}(L^{24}L^{13}+L^{34}L^{12})+(q^{2}+q^{-2})L^{14}%
L^{14}\nonumber\\
&  -\;\lambda(L^{14}L^{23}+L^{23}L^{14}),\nonumber
\end{align}
and the second one by
\begin{align}
C_{2}  &  =\varepsilon_{ijkl}L^{ij}L^{kl}=q^{2}\lambda_{+}^{2}(L^{14}%
L^{23}+L^{23}L^{14})+q^{2}\lambda_{+}^{2}(L^{12}L^{34}-L^{13}L^{24})\\
&  +\;q^{4}\lambda_{+}^{2}(L^{34}L^{12}-L^{24}L^{13})-q^{2}\lambda\lambda
_{+}^{2}L^{14}L^{14},\nonumber
\end{align}
where $g_{ik}$ and $\varepsilon_{ijkl}$ denote respectively quantum metric and
q-deformed epsilon-tensor (for its definition see Ref. \cite{Fiore})
corresponding to $U_{q}(so_{4})$. Again, we would like to specify the Casimir
operators for the different representations. In doing so, we get the following:

\begin{enumerate}
\item[a)] (operator representation)
\begin{align}
C_{1}  &  =2q^{-2}(X\circ X)(\partial\circ\partial)+2q^{2}(X\circ
\partial)(X\circ\partial)\\
&  +\;2q^{-1}\lambda_{+}X\circ\partial,\nonumber\\
C_{2}  &  =0,\nonumber
\end{align}
with $U\circ V\equiv g_{AB}U^{A}V^{B}$,

\item[b)] (spinor representation)
\begin{align}
C_{1}  &  =[[3]]_{q^{-4}}\mbox{1 \kern-.59em {\rm l}}_{2\times2},\\
C_{2}  &  =q[[3]]_{q^{4}}\lambda_{+}\mbox{1 \kern-.59em {\rm l}}_{2\times
2},\nonumber
\end{align}

\item[c)] (vector representation)
\begin{align}
C_{1}  &  =2[[3]]_{q^{-4}}\mbox{1 \kern-.59em {\rm l}}_{3\times3},\\
C_{2}  &  =0.\nonumber
\end{align}

\end{enumerate}

\section{Quantum Lie algebra of Lorentz transformations}

In this section we deal with the quantum Lie algebra of Lorentz
transformations. Everything so far applies to this case, which from a physical
point of view is the most interesting one we consider in this article.

\subsection{Representation of Lorentz generators within q-deformed
differential calculus}

In complete analogy to the previous section we start with a realization of the
$V^{\mu\nu}$ given by \cite{LWW97}
\begin{equation}
V^{\mu\nu}\equiv\Lambda^{1/2}(P_{A})^{\mu\nu}{}_{\rho\sigma}\,X^{\rho}%
\hat{\partial}^{\sigma}=-q^{-2}\Lambda^{1/2}(P_{A})^{\mu\nu}{}_{\rho\sigma
}\,\hat{\partial}^{\rho}X^{\sigma},
\end{equation}
or equivalently%
\begin{equation}
V^{\mu\nu}\equiv q^{-2}\Lambda^{-1/2}(P_{A})^{\mu\nu}{}_{\rho\sigma}\,X^{\rho
}\partial^{\sigma}=-\Lambda^{-1/2}(P_{A})^{\mu\nu}{}_{\rho\sigma}%
\,\partial^{\rho}X^{\sigma},
\end{equation}
where $\Lambda$ is a scaling operator being subject to
\begin{equation}
\Lambda X^{\mu}=q^{-2}X^{\mu}\Lambda,\quad\Lambda\partial^{\mu}=q^{2}%
\partial^{\mu}\Lambda,
\end{equation}
and $P_{A}$ denotes the antisymmetrizer for q-deformed Lorentz symmetry. More
explicitly, we have for example
\begin{align}
V^{+3}  &  =2q^{2}\lambda_{+}^{-2}X^{+}\hat{\partial}^{3}-2\lambda_{+}%
^{-2}X^{3}\hat{\partial}^{+}\\
&  -\;q\lambda\lambda_{+}^{-2}(X^{+}\hat{\partial}^{0}-X^{0}\hat{\partial}%
^{+}),\nonumber\\
V^{+0}  &  =2\lambda_{+}^{-2}X^{+}\hat{\partial}^{0}-q^{-2}(q^{4}%
+1)\lambda_{+}^{-2}X^{0}\hat{\partial}^{+}\nonumber\\
&  +\;\lambda\lambda_{+}^{-2}(qX^{+}\hat{\partial}^{3}-q^{-1}X^{3}%
\hat{\partial}^{+}),\nonumber\\
V^{+-}  &  =2\lambda_{+}^{-2}(\lambda^{2}+1)X^{+}\hat{\partial}^{-}%
-2\lambda_{+}^{-2}X^{-}\hat{\partial}^{+}\nonumber\\
&  -\;\lambda\lambda_{+}^{-2}(X^{3}\hat{\partial}^{0}+X^{0}\hat{\partial}%
^{3}),\nonumber\\
V^{30}  &  =2\lambda_{+}^{-2}X^{3}\hat{\partial}^{0}-q^{-2}(q^{4}%
+1)\lambda_{+}^{-2}X^{0}\hat{\partial}^{3}\nonumber\\
&  +\;\lambda\lambda_{+}^{-2}(\lambda^{2}+1)(X^{+}\hat{\partial}^{-}-X^{-}%
\hat{\partial}^{+}),\nonumber\\
V^{3-}  &  =2q^{2}\lambda_{+}^{-2}X^{3}\hat{\partial}^{-}-2\lambda_{+}%
^{-2}X^{-}\hat{\partial}^{3}\nonumber\\
&  -\;q\lambda\lambda_{+}^{-2}(X^{0}\hat{\partial}^{-}+X^{-}\hat{\partial}%
^{0}),\nonumber\\
V^{0-}  &  =2\lambda_{+}^{-2}X^{0}\hat{\partial}^{-}-q^{-2}(q^{4}%
+1)\lambda_{+}^{-2}X^{-}\hat{\partial}^{0}\nonumber\\
&  -\;\lambda\lambda_{+}^{-2}(qX^{3}\hat{\partial}^{-}-q^{-1}X^{-}%
\hat{\partial}^{3}).\nonumber
\end{align}
The antisymmetry of the $V^{\mu\nu}$ implies that we have six independent
components for which we can choose
\begin{equation}
V^{+3},V^{+0},V^{+-},V^{30},V^{3-},V^{0-}, \label{inGenLor}%
\end{equation}
since the remaining ones are related to them by the relations%
\begin{gather}
V^{++}=V^{00}=V^{--}=0,\\
V^{33}=\lambda V^{+-},\quad V^{-+}=-V^{+-},\nonumber\\
V^{\pm3}=-q^{\pm2}V^{3\pm},\nonumber\\
V^{\pm0}=\mp q^{\pm1}\lambda V^{3\pm}-V^{0\pm},\nonumber\\
V^{03}=\lambda V^{+-}-V^{30}.\nonumber
\end{gather}

In Ref. \cite{LWW97} it was shown that the $V^{\mu\nu}$ together with two
additional generators, denoted in the following by $U^{1},U^{2}$, span
q-deformed Lorentz algebra. However, in what follows it is convenient to
introduce another set of generators which are related to the $V^{\mu\nu}$ by
\begin{align}
R^{+}  &  =q^{-1}\lambda_{+}^{-1}(V^{+0}-V^{+3}),\label{RS-Form}\\
R^{3}  &  =q^{-1}\lambda_{+}^{-1}(V^{30}-qV^{+-}),\nonumber\\
R^{-}  &  =q^{-1}\lambda_{+}^{-1}(q^{-2}V^{3-}+V^{0-}),\nonumber\\[0.1in]
S^{+}  &  =q^{-1}\lambda_{+}^{-1}(V^{+0}+q^{-2}V^{+3}),\label{RS-Form2}\\
S^{3}  &  =q^{-1}\lambda_{+}^{-1}(V^{30}+q^{-1}V^{+-}),\nonumber\\
S^{-}  &  =q^{-1}\lambda_{+}^{-1}(V^{3-}-V^{0-}).\nonumber
\end{align}
In terms of the generators $R^{A}$ and $S^{A}$ the relations of q-deformed
Lorentz algebra take the rather compact form
\begin{align}
{\epsilon_{CB}}^{A}R^{B}R^{C}  &  =q^{-4}\lambda_{+}^{-1}U^{1}R^{A}%
,\label{RelRS}\\
{\epsilon_{CB}}^{A}S^{B}S^{C}  &  =-q^{-4}\lambda_{+}^{-1}U^{2}S^{A}%
,\nonumber\\
R^{A}S^{B}  &  =q^{2}(\hat{R}_{SO_{q}(3)})_{CD}^{AB}\,S^{C}R^{D},\nonumber
\end{align}
while $U^{1}$ and $U^{2}$ are both central in the algebra. ${\epsilon_{CB}%
}^{A}$ in (\ref{RelRS}) denotes the epsilon tensor from Sec. \ref{QuLie3dim}
and $\hat{R}_{SO_{q}(3)}$ stands for the vector representation of the
universal R-matrix of $SO_{q}(3).$ For more details we refer the reader to
\cite{LWW97}.

\subsection{Hopf structure of q-deformed Lorentz algebra and corresponding
q-commutators}

To write down q-commutators with Lorentz generators $V^{\mu\nu}$ we need to
know their Hopf structure. From Ref. \cite{Rohr99}, however, we know the
coproduct for the generators $R^{A}$ and $S^{A},$ $A\in\{+,3,-\}.$ By solving
(\ref{RS-Form}) and (\ref{RS-Form2}) for the independent generators given in
(\ref{inGenLor}) and making use of the algebra homomorphism property of the
coproduct we are able to find expressions for the coproducts of the $V^{\mu
\nu}$.

For compactness, we introduce
\begin{equation}
\rho\equiv q^{2}\lambda\lambda_{+}R^{3}+U^{1},\quad\sigma\equiv q^{2}%
\lambda\lambda_{+}S^{3}-U^{2},
\end{equation}
and the generators of an $U_{q}(su_{2})$-subalgebra, given by \cite{Rohr99}%
\begin{align}
L^{A}  &  \equiv-q^{2}\lambda_{+}(U^{1}S^{A}-U^{2}R^{A}+q^{2}\lambda
\lambda_{+}\,\epsilon_{CB}{}^{A}R^{B}S^{C}),\\
\tau^{-1/2}  &  \equiv U^{1}U^{2}-q^{4}\lambda\lambda_{+}\,g_{AB}\,R^{A}%
S^{B}+\lambda L^{3},\nonumber
\end{align}
where $g_{AB}$ is the quantum metric of Sec. \ref{QuLie3dim}.\ In terms of the
new quantities the wanted coproducts read as
\begin{align}
\Delta(V^{+3})  &  =-\;q^{2}\sigma\otimes R^{+}+q^{2}R^{+}\otimes\rho\\
&  -\;q^{2}\tau^{1/2}R^{+}\otimes\tau^{1/2}\rho+q^{2}\tau^{1/2}\sigma
\otimes\tau^{1/2}R^{+}\nonumber\\
&  -\;q^3\lambda_{+}^{-1}\tau^{1/2}\sigma L^{+}\otimes\sigma-q^3\lambda_{+}^{-1}%
\tau^{1/2}\sigma\otimes\tau^{1/2}\sigma L^{+}\nonumber\\
&  +\;q^{7}\lambda^{2}\lambda_{+}^{2}\tau^{1/2}R^{+}\otimes\tau^{1/2}%
S^{-}L^{+}+q^{9}\lambda^{2}\lambda_{+}^{2}\tau^{1/2}R^{+}L^{+}\otimes
S^{-},\nonumber\\
\Delta(V^{+0})  &  =\sigma\otimes R^{+}-R^{+}\otimes\rho\nonumber\\
&  -\;q^{2}\tau^{1/2}R^{+}\otimes\tau^{1/2}\rho+q^{2}\tau^{1/2}\sigma
\otimes\tau^{1/2}R^{+}\nonumber\\
&  -\;q^3\lambda_{+}^{-1}\tau^{1/2}\sigma L^{+}\otimes\sigma-q^3\lambda_{+}^{-1}%
\tau^{1/2}\sigma\otimes\tau^{1/2}\sigma L^{+}\nonumber\\
&  +\;q^{7}\lambda^{2}\lambda_{+}^{2}\tau^{1/2}R^{+}\otimes\tau^{1/2}%
S^{-}L^{+}+q^{9}\lambda^{2}\lambda_{+}^{2}\tau^{1/2}R^{+}L^{+}\otimes
S^{-},\nonumber\\
\Delta(V^{+-})  &  =qR^{3}\otimes\rho-q\tau^{1/2}\sigma\otimes R^{3}%
\nonumber\\
&  +\;qS^{3}\otimes\sigma-q\tau^{1/2}\rho\otimes S^{3}\nonumber\\
&  -\;q^{4}\lambda\lambda_{+}S^{-}\otimes R^{+}-q^{2}\lambda\lambda
_{+}R^{+}\otimes S^{-}\nonumber\\
&  +\;q^{4}\lambda\lambda_{+}\tau^{1/2}R^{+}\otimes R^{-}+q^3\lambda\tau
^{1/2}\sigma L^{-}\otimes R^{+}\nonumber\\
&  +\;q^{2}\lambda\lambda_{+}\tau^{1/2}S^{-}\otimes S^{+}+q^{5}\lambda
\tau^{1/2}\rho L^{+}\otimes S^{-},\nonumber\\
\Delta(V^{30})  &  =-\;R^{3}\otimes\rho+\tau^{1/2}\sigma\otimes R^{3}%
\nonumber\\
&  +\;q^{2}S^{3}\otimes\sigma-q^{2}\tau^{1/2}\rho\otimes S^{3}\nonumber\\
&  +\;q^{3}\lambda\lambda_{+}S^{-}\otimes R^{+}-q^{3}\lambda\lambda
_{+}R^{+}\otimes S^{-}\nonumber\\
&  -\;q^{3}\lambda\lambda_{+}\tau^{1/2}R^{+}\otimes R^{-}-q^{2}\lambda
\tau^{1/2}\sigma L^{-}\otimes R^{+}\nonumber\\
&  +\;q^{3}\lambda\lambda_{+}\tau^{1/2}S^{-}\otimes S^{+}+q^{6}\lambda
\tau^{1/2}\rho L^{+}\otimes S^{-},\nonumber\\
\Delta(V^{3-})  &  =q^{2}S^{-}\otimes\sigma-q^{2}\rho\otimes S^{-}\nonumber\\
&  -\;q^{2}\tau^{1/2}S^{-}\otimes\tau^{1/2}\sigma+q^{2}\tau^{1/2}\rho
\otimes\tau^{1/2}S^{-}\nonumber\\
&  -\;q^3\lambda_{+}^{-1}\tau^{1/2}\rho L^{-}\otimes\rho-q^3\lambda_{+}^{-1}%
\tau^{1/2}\rho\otimes\tau^{1/2}\rho L^{-}\nonumber\\
&  +\;q^{5}\lambda^{2}\lambda_{+}^{2}\tau^{1/2}S^{-}L^{-}\otimes R^{+}%
+q^{7}\lambda^{2}\lambda_{+}^{2}\tau^{1/2}S^{-}\otimes\tau^{1/2}R^{+}%
L^{-},\nonumber\\
\Delta(V^{0-})  &  =-\;S^{-}\otimes\sigma+\rho\otimes S^{-}\nonumber\\
&  -\;q^{2}\tau^{1/2}S^{-}\otimes\tau^{1/2}\sigma+q^{2}\tau^{1/2}\rho
\otimes\tau^{1/2}S^{-}\nonumber\\
&  -\;q^3\lambda_{+}^{-1}\tau^{1/2}\rho L^{-}\otimes\rho-q^3\lambda_{+}^{-1}%
\tau^{1/2}\rho\otimes\tau^{1/2}\rho L^{-}\nonumber\\
&  +\;q^{5}\lambda^{2}\lambda_{+}^{2}\tau^{1/2}S^{-}L^{-}\otimes R^{+}%
+q^{7}\lambda^{2}\lambda_{+}^{2}\tau^{1/2}S^{-}\otimes\tau^{1/2}R^{+}%
L^{-}.\nonumber
\end{align}
For writing down q-commutators we also need the following antipodes and their
inverses:%
\begin{align}
S(R^{+})  &  =q^{2}S^{-1}(R^{+})=-q^{2}\tau^{1/2}R^{+},\\
S(R^{3})  &  =S^{-1}(R^{3})=-q^{-2}\lambda^{-1}\lambda_{+}^{-1}(U^{1}%
+\tau^{1/2})\sigma,\nonumber\\
S(R^{-})  &  =q^{-2}S^{-1}(R^{-})=-S^{-}-q^{-1}\lambda_{+}^{-1}\tau^{1/2}%
L^{-}\sigma,\nonumber\\[0.16in]
S(S^{+})  &  =q^{2}S^{-1}(S^{+})=-R^{+}-q^{3}\lambda_{+}^{-1}\tau^{1/2}%
L^{+}\rho,\\
S(S^{3})  &  =S^{-1}(S^{3})=q^{-2}\lambda^{-1}\lambda_{+}^{-1}(U^{2}%
-\tau^{1/2}\rho),\nonumber\\
S(S^{-})  &  =q^{-2}S^{-1}(S^{-})=-q^{-2}\tau^{1/2}S^{-},\nonumber\\[0.16in]
S(U^{i})  &  =S^{-1}(U^{i})=U^{i},\quad i=1,2,\\[0.16in]
S(\rho)  &  =S^{-1}(\rho)=-\tau^{1/2}\sigma,\\
S(\sigma)  &  =S^{-1}(\sigma)=-\tau^{1/2}\rho.\nonumber
\end{align}

After these preparations we are ready to write down q-commutators. For their
left versions we have found
\begin{align}
{[}V^{+3},V{]}_{q}  &  =q^3\lambda_{+}^{-1}\tau^{1/2}\sigma(L^{+}V-VL^{+}%
)\tau^{1/2}\rho\\
&  +\;q^{7}\lambda^{2}\lambda_{+}^{2}\tau^{1/2}R^{+}(VL^{+}-L^{+}V)\tau
^{1/2}S^{-}\nonumber\\
&  +\;q^{2}(q^{2}\sigma V\tau^{1/2}R^{+}-\tau^{1/2}\sigma VR^{+})\nonumber\\
&
+\;q^{2}(\tau^{1/2}R^{+}V\sigma-R^{+}V\tau^{1/2}\sigma),\nonumber\\[0.1in]
{[}V^{+0},V{]}_{q}  &  =q^3\lambda_{+}^{-1}\tau^{1/2}\sigma(L^{+}V-VL^{+}%
)\tau^{1/2}\rho\\
&  +\;q^{7}\lambda^{2}\lambda_{+}^{2}\tau^{1/2}R^{+}(VL^{+}-L^{+}V)\tau
^{1/2}S^{-}\nonumber\\
&  -\;q^{2}(\sigma V\tau^{1/2}R^{+}+\tau^{1/2}\sigma VR^{+})\nonumber\\
&
+\;q^{2}\tau^{1/2}R^{+}V\sigma+R^{+}V\tau^{1/2}\sigma,\nonumber\\[0.1in]
{[}V^{+-},V{]}_{q}  &  =\;q(R^{3}V\tau^{1/2}\sigma+S^{3}V\tau^{1/2}%
\rho)\\
&  +\;q^{-1}\lambda^{-1}\lambda_{+}^{-1}\tau^{1/2}(\sigma V(U^{1}+\tau
^{1/2}\sigma)-\rho V(U^{2}-\tau^{1/2}\rho))\nonumber\\
&  -\;q^3\lambda\tau^{1/2}(\rho L^{+}V\tau^{1/2}S^{-}+q^{2}S^{-}V\tau^{1/2}%
L^{+}\rho)\nonumber\\
&  -\;q^3\lambda\tau^{1/2}(q^{2}\sigma L^{-}V\tau^{1/2}R^{+}+R^{+}V\tau
^{1/2}L^{-}\sigma)\nonumber\\
&  +\;q^{4}\lambda\lambda_{+}(q^{2}S^{-}V\tau^{1/2}R^{+}-\tau^{1/2}R^{+}%
VS^{-})\nonumber\\
&  +\;\lambda\lambda_{+}(R^{+}V\tau^{1/2}S^{-}-q^{2}\tau^{1/2}S^{-}%
VR^{+}),\nonumber\\[0.1in]
{[}V^{30},V{]}_{q}  &  =R^{3}V\tau^{1/2}\sigma-q^{2}S^{3}V\tau^{1/2}\rho\\
&  -\;q^{-2}\lambda^{-1}\lambda_{+}^{-1}\tau^{1/2}(\sigma V(U^{1}+\tau
^{1/2}\sigma))+q^{2}\rho V(U^{2}-\tau^{1/2}\rho)\nonumber\\
&  -\;q^4\lambda\tau^{1/2}(\rho L^{+}V\tau^{1/2}S^{-}+q^{2}S^{-}V\tau^{1/2}%
L^{+}\rho)\nonumber\\
&  +\;q^{2}\lambda\tau^{1/2}(q^{2}\sigma L^{-}V\tau^{1/2}R^{+}+R^{+}%
V\tau^{1/2}L^{-}\sigma)\nonumber\\
&  +\;q^{3}\lambda\lambda_{+}(q^{2}S^{-}V\tau^{1/2}R^{+}-\tau^{1/2}R^{+}%
VS^{-})\nonumber\\
&  +\;q\lambda\lambda_{+}(R^{+}V\tau^{1/2}S^{-}-q^{2}\tau^{1/2}S^{-}%
VR^{+}),\nonumber\\[0.1in]
{[}V^{3-},V{]}_{q}  &  =\rho V\tau^{1/2}S^{-}-q^{2}\tau^{1/2}\rho VS^{-}\\
&  +\;q^{2}(\tau^{1/2}S^{-}V\rho-S^{-}V\tau^{1/2}\rho)\nonumber\\
&  -q^3\;\lambda_{+}^{-1}\tau^{1/2}\rho(L^{-}V-VL^{-})\tau^{1/2}\sigma\nonumber\\
&  \;q^{7}\lambda^{2}\lambda_{+}^{2}\tau^{1/2}S^{-}(VL^{-}-L^{-}V)\tau
^{1/2}R^{+},\nonumber\\[0.1in]
{[}V^{0-},V{]}_{q}  &  =-\;q^{-2}\rho V\tau^{1/2}S^{-}-q^{2}\tau^{1/2}\rho
VS^{-}\\
&  +\;q^{2}\tau^{1/2}S^{-}V\rho+S^{-}V\tau^{1/2}\rho\nonumber\\
&  -\;q^3\lambda_{+}^{-1}\tau^{1/2}\rho(L^{-}V-VL^{-})\tau^{1/2}\sigma\nonumber\\
&  \;q^{7}\lambda^{2}\lambda_{+}^{2}\tau^{1/2}S^{-}(VL^{-}-L^{-}V)\tau
^{1/2}R^{+},\nonumber
\end{align}
and likewise for the corresponding right versions,
\begin{align}
{[}V,V^{+3}{]}_{q}  &  =q^{2}\sigma V\tau^{1/2}R^{+}-q^{2}\sigma\tau
^{1/2}VR^{+}\\
&  -\;R^{+}V\tau^{1/2}\sigma+R^{+}\tau^{1/2}V\sigma\nonumber\\
&  -q^3\;\lambda_{+}^{-1}(\rho V\tau^{1/2}\sigma L^{+}-L^{+}\rho V\tau
^{1/2}\sigma)\nonumber\\
&  +\;q^{9}\lambda^{2}\lambda_{+}^{2}\tau^{1/2}(L^{+}S^{-}V\tau^{1/2}%
R^{+}-q^2 S^{-}V\tau^{1/2}R^{+}L^{+}),\nonumber\\[0.1in]
{[}V,V^{+0}{]}_{q}  &  =q^{2}\sigma V\tau^{1/2}R^{+}+\sigma\tau^{1/2}VR^{+}\\
&  -\;R^{+}V\tau^{1/2}\sigma-q^{-2}R^{+}\tau^{1/2}V\sigma\nonumber\\
&  -q^3\;\lambda_{+}^{-1}(\rho V\tau^{1/2}\sigma L^{+}-L^{+}\rho V\tau
^{1/2}\sigma)\nonumber\\
&  +\;q^{9}\lambda^{2}\lambda_{+}^{2}\tau^{1/2}(L^{+}S^{-}V\tau^{1/2}%
R^{+}-q^2 S^{-}V\tau^{1/2}R^{+}L^{+}),\nonumber\\[0.1in]
{[}V,V^{+-}{]}_{q}  &  =-\;q(\rho\tau^{1/2}VS^{3}+\sigma\tau^{1/2}VR^{3})\\
&  +\;q^{-1}\lambda^{-1}\lambda_{+}^{-1}((-U^{2}+\rho\tau^{1/2})V\tau^{1/2}%
\rho+(U^{1}+\sigma\tau^{1/2})V\tau^{1/2}\sigma)\nonumber\\
&  -\;q\lambda(L^{+}\rho\tau^{1/2} V\tau^{1/2}S^{-}+q^{6}S^{-}\tau^{1/2}%
V\tau^{1/2}\rho L^{+})\nonumber\\
&  -\;q\lambda(q^{4}\tau^{1/2}L^{-}\sigma V\tau^{1/2}R^{+}+R^{+}%
\tau^{1/2}V\tau^{1/2}\sigma L^{-})\nonumber\\
&  +\;\lambda\lambda_{+}(q^{4}S^{-}\tau^{1/2}VR^{+}-R^{+}V\tau^{1/2}%
S^{-})\nonumber\\
&  +\;q^{2}\lambda\lambda_{+}(R^{+}\tau^{1/2}VS^{-}-q^{4}S^{-}V\tau^{1/2}%
R^{+}),\nonumber\\[0.1in]
{[}V,V^{30}{]}_{q}  &  =\sigma\tau^{1/2}VR^{3}-q^{2}\rho\tau^{1/2}VS^{3}\\
&  -\;q^{-2}\lambda^{-1}\lambda_{+}^{-1}((U^{1}+\sigma\tau^{1/2})V\tau
^{1/2}\sigma+q^{2}(U^{2}-\rho\tau^{1/2})V\tau^{1/2}\rho)\nonumber\\
&  +\;\lambda(R^{+}\tau^{1/2}V\tau^{1/2}\sigma L^{-}+q^4\tau^{1/2}%
L^{-}\sigma V\tau^{1/2}R^{+})\nonumber\\
&  -\;q^{2}\lambda(L^{+}\rho\tau^{1/2}V\tau^{1/2}S^{-}-q^{6}S^{-}\tau
^{1/2}V\tau^{1/2}\rho L^{+})\nonumber\\
&  -\;q\lambda\lambda_{+}(R^{+}V\tau^{1/2}S^{-}+R^{+}\tau^{1/2}VS^{-}%
)\nonumber\\
&  +\;q^{5}\lambda\lambda_{+}(S^{-}V\tau^{1/2}R^{+}+S^{-}\tau^{1/2}%
VR^{+}),\nonumber\\[0.1in]
{[}V,V^{3-}{]}_{q}  &  =-\;q^{2}\rho\tau^{1/2}VS^{-}+q^{2}\rho V\tau
^{1/2}S^{-}\\
&  -\;q^{4}S^{-}V\tau^{1/2}\rho+q^{4}S^{-}\tau^{1/2}V\rho\nonumber\\
&  +\;q^3\lambda_{+}^{-1}(\sigma\tau^{1/2}V\tau^{1/2}\rho L^{-}-\tau^{1/2}%
L^{-}\sigma V\tau^{1/2}\rho)\nonumber\\
&  +\;q^3\lambda^{2}\lambda_{+}^{2}(q^2\tau^{1/2}L^{-}R^{+}V\tau^{1/2}S^{-}%
-R^{+}\tau^{1/2}V\tau^{1/2}S^{-}L^{-}),\nonumber\\[0.1in]
{[}V,V^{0-}{]}_{q}  &  =\rho\tau^{1/2}VS^{-}+q^{2}\rho V\tau^{1/2}S^{-}\\
&  -\;q^{4}S^{-}V\tau^{1/2}\rho-q^{2}S^{-}\tau^{1/2}V\rho\nonumber\\
&  +\;q^3\lambda_{+}^{-1}(\sigma\tau^{1/2}V\tau^{1/2}\rho L^{-}-\tau^{1/2}%
L^{-}\sigma V\tau^{1/2}\rho)\nonumber\\
&  -\;q^3\lambda^{2}\lambda_{+}^{2}(q^2\tau^{1/2}L^{-}R^{+}V\tau^{1/2}S^{-}%
-R^{+}\tau^{1/2}V\tau^{1/2}S^{-}L^{-}).\nonumber
\end{align}
\vspace{0.05cm}

\subsection{Matrix representations of q-deformed Lorentz algebra and
commutation relations with tensor operators}

Let us now consider spin representations of q-deformed Lorentz algebra. From a
physical point of view q-deformed analogs of the two spinor representations
$(1/2,0)$ and $(0,1/2)$ as well as the vector representation $(1/2,1/2)$ are
the most interesting cases \cite{CSSW90, SWZ91,OSWZ92,LSW94}. For a systematic
treatment of finite dimensional representations of q-Lorentz algebra we refer
for example to \cite{Bloh01}. As in the previous sections, we begin with
spinor representations, for which we have
\begin{align}
V^{\mu\nu}\rhd\theta^{\,\alpha}  &  =\left(  \sigma^{\mu\nu}\right)  ^{\alpha
}{}_{\beta}\;\theta^{\beta}, & V^{\mu\nu}\rhd\bar{\theta}^{\,\alpha}  &
=\left(  \bar{\sigma}^{\mu\nu}\right)  ^{\alpha}{}_{\beta}\;\bar{\theta
}^{\beta},\\
\theta_{\alpha}\lhd V^{\mu\nu}  &  =\theta_{\beta}\,\left(  \sigma^{\mu\nu
}\right)  ^{\beta}{}_{\alpha}\;, & \bar{\theta}_{\alpha}\lhd V^{\mu\nu}  &
=\bar{\theta}_{\beta}\,\left(  \bar{\sigma}^{\mu\nu}\right)  ^{\beta}%
{}_{\alpha}\;,\nonumber
\end{align}
and the spin matrices are given by
\begin{align}
(\sigma^{+3})^{\alpha}{}_{\beta}  &  =q^{2}(\sigma^{+0})^{\alpha}{}_{\beta
}=-q^{1/2}\lambda_{+}^{-3/2}\left(
\begin{array}
[c]{cc}%
0 & 0\\
1 & 0
\end{array}
\right)  ,\\
(\sigma^{+-})^{\alpha}{}_{\beta}  &  =-q(\sigma^{30})^{\alpha}{}_{\beta
}=q^{-1}\lambda_{+}^{-2}\left(
\begin{array}
[c]{cc}%
-q & 0\\
0 & q^{-1}%
\end{array}
\right)  ,\nonumber\\
(\sigma^{3-})^{\alpha}{}_{\beta}  &  =(\sigma^{0-})^{\alpha}{}_{\beta
}=q^{-1/2}\lambda_{+}^{-3/2}\left(
\begin{array}
[c]{cc}%
0 & 1\\
0 & 0
\end{array}
\right)  ,\nonumber
\end{align}
in connection with
\begin{align}
(\bar{\sigma}^{+3})^{\alpha}{}_{\beta}  &  =(\sigma^{+3})^{\alpha}{}_{\beta},
& (\bar{\sigma}^{+0})^{\alpha}{}_{\beta}  &  =-q^{2}(\sigma^{+0})^{\alpha}%
{}_{\beta},\\
(\bar{\sigma}^{+-})^{\alpha}{}_{\beta}  &  =(\sigma^{+-})^{\alpha}{}_{\beta},
& (\bar{\sigma}^{30})^{\alpha}{}_{\beta}  &  =-q^{2}(\sigma^{30})^{\alpha}%
{}_{\beta},\nonumber\\
(\bar{\sigma}^{3-})^{\alpha}{}_{\beta}  &  =(\sigma^{3-})^{\alpha}{}_{\beta},
& (\bar{\sigma}^{0-})^{\alpha}{}_{\beta}  &  =-q^{2}(\sigma^{0-})^{\alpha}%
{}_{\beta}.\nonumber
\end{align}
The spinor representations of the generators $U^{1}$, $U^{2},$ and $\sigma$
are given by diagonal matrices which fulfill the identities%
\begin{equation}
-q^{2}(q^{2}+q^{-2})^{-1}\lambda_{+}{(U^{1})^{\alpha}}_{\beta}=-{(U^{2}%
)^{\alpha}}_{\beta}=({\sigma)^{\alpha}}_{\beta}={\delta^{\alpha}}_{\beta}.
\end{equation}
Furthermore, for $\rho$ and the generators of the $U_{q}(su_{2})$-subalgebra
we have
\begin{gather}
{(L^{+})^{\alpha}}_{\beta}=-q^{1/2}\lambda_{+}^{-1/2}%
\begin{pmatrix}
0 & 0\\
1 & 0
\end{pmatrix}
,\quad{(L^{-})^{\alpha}}_{\beta}=-q^{-1/2}\lambda_{+}^{-1/2}%
\begin{pmatrix}
0 & 1\\
0 & 0
\end{pmatrix}
,\\
{(\tau^{-1/2})^{\alpha}}_{\beta}=-({\rho)^{\alpha}}_{\beta}=%
\begin{pmatrix}
q^{-1} & 0\\
0 & q
\end{pmatrix}
.\nonumber
\end{gather}

What we have done so far, enables us to write out the commutation relations
between Lorentz generators and components of a spinor operator. In general,
they are equivalent to%
\begin{align}
{[}V^{\mu\nu},\theta^{\,\alpha}{]}_{q}  &  =(\sigma^{\mu\nu})^{\alpha}%
{}_{\beta}\,\theta^{\beta},\quad{[}V^{\mu\nu},\bar{\theta}^{\,\alpha}{]}%
_{q}=(\bar{\sigma}^{\mu\nu})^{\alpha}{}_{\beta}\,\bar{\theta}^{\beta
},\label{Vtheta}\\
{[}\theta_{\alpha},V^{\mu\nu}{]}_{q}  &  =\theta_{\beta}\,(\sigma^{\mu\nu
})^{\beta}{}_{\alpha},\quad{[}\bar{\theta}_{\alpha},V^{\mu\nu}{]}_{q}%
=\bar{\theta}_{\beta}\,(\bar{\sigma}^{\mu\nu})^{\beta}{}_{\alpha}.\nonumber
\end{align}
Unfortunately, it is rather difficult to derive from (\ref{Vtheta}) the
explicit form of the commutation relations between Lorentz generators and
spinors. Thus, we proceed differently and apply the identities%
\begin{align}
V^{\mu\nu}a  &  =((V^{\mu\nu})_{(1)}\triangleright a)(V^{\mu\nu}%
)_{(2)},\label{VerLoGenSp}\\
aV^{\mu\nu}  &  =(V^{\mu\nu})_{(2)}(a\triangleleft(V^{\mu\nu})_{(1)}%
),\nonumber
\end{align}
which result from Eqs. (\ref{LefCrosPro}) and (\ref{RigCrosPro}). This way the
first relation in (\ref{VerLoGenSp}) implies
\begin{align}
V^{+3}\theta^{1}  &  =(q+1)\lambda_{+}^{-1}\theta^{1}V^{+3}+(q^{2}%
-q)\lambda_{+}^{-1}\theta^{1}V^{+0}\\
&  -\;q^{1/2}\lambda\lambda_{+}^{-1/2}\theta^{2}(V^{+-}+qV^{30})+q^{1/2}%
\lambda_{+}^{-3/2}\theta^{2}\rho,\nonumber\\
V^{+3}\theta^{2}  &  =(q+q^{-2})\lambda_{+}^{-1}\theta^{2}V^{+3}%
+(1-q)\lambda_{+}^{-1}\theta^{2}V^{+0},\nonumber\\[0.16in]
V^{+0}\theta^{1}  &  =(q^{2}+q^{-1})\lambda_{+}^{-1}\theta^{1}V^{+0}%
+(1-q^{-1})\lambda_{+}^{-1}\theta^{1}V^{+3}\\
&  -\;q^{1/2}\lambda\lambda_{+}^{-1/2}\theta^{2}(V^{+-}+qV^{30})-q^{-3/2}%
\lambda_{+}^{-3/2}\theta^{2}\rho,\nonumber\\
V^{+0}\theta^{2}  &  =(1+q^{-1})\lambda_{+}^{-1}\theta^{2}V^{+0}%
+(q^{-2}-q^{-1})\lambda_{+}^{-1}\theta^{2}V^{+3},\nonumber\\[0.16in]
V^{+-}\theta^{1}  &  =q^{-1}(2q^{2}+\lambda_{+})\lambda_{+}^{-2}\theta
^{1}V^{+-}+q^{-1}(q-1)^{2}\lambda_{+}^{-2}\theta^{1}V^{30}\\
&  +\;q^{1/2}(\lambda_{+}-1)\lambda\lambda_{+}^{-3/2}\theta^{2}V^{0-}%
\nonumber\\
&  -\;q^{-3/2}(q^{2}\lambda_{+}+1)\lambda\lambda_{+}^{-3/2}\theta^{2}%
V^{3-}+\lambda_{+}^{-2}\theta^{1}U^{1},\nonumber\\
V^{+-}\theta^{2}  &  =q^{-1}(2q^{2}+\lambda_{+})\lambda_{+}^{-2}\theta
^{2}V^{+-}+q^{-1}(q-1)^{2}\lambda_{+}^{-2}\theta^{2}V^{30}\nonumber\\
&  +\;q^{-1/2}\lambda\lambda_{+}^{-3/2}\theta^{1}(V^{+0}-V^{+3})-q^{-2}%
\lambda_{+}^{-2}\theta^{2}U^{1},\nonumber\\[0.16in]
V^{30}\theta^{1}  &  =q^{-1}(q^{2}\lambda_{+}+2)\lambda_{+}^{-2}\theta
^{1}V^{30}+q^{-1}(q-1)^{2}\lambda_{+}^{-2}\theta^{1}V^{+-}\\
&  +\;q^{-1/2}(1+q^{2}\lambda_{+})\lambda\lambda_{+}^{-3/2}\theta
^{2}V^{0-}\nonumber\\
&  +\;q^{-5/2}(1-q^{4}\lambda_{+})\lambda\lambda_{+}^{-3/2}\theta
^{2}V^{3-}-q^{-1}\lambda_{+}^{-2}\theta^{1}U^{1},\nonumber\\
V^{30}\theta^{2}  &  =q^{-1}(q^{2}\lambda_{+}+2)\lambda_{+}^{-2}\theta
^{2}V^{30}+q^{-1}(q-1)^{2}\lambda_{+}^{-2}\theta^{2}V^{+-}\nonumber\\
&  +\;q^{-3/2}\lambda\lambda_{+}^{-3/2}\theta^{1}(V^{+0}-V^{+3})+q^{-3}%
\lambda_{+}^{-2}\theta^{2}U^{1},\nonumber\\[0.16in]
V^{3-}\theta^{1}  &  =(q^{-1}+1)\lambda_{+}^{-1}\theta^{1}V^{3-}%
+(q-1)\lambda_{+}^{-1}\theta^{2}V^{0-},\\
V^{3-}\theta^{2}  &  =(q^{2}+q^{-1})\lambda_{+}^{-1}\theta^{2}V^{3-}%
-(q^{2}-q)\lambda_{+}^{-1}\theta^{1}V^{0-}\nonumber\\
&  -\;q^{-1/2}\lambda_{+}^{-3/2}\theta^{1}\rho,\nonumber\\[0.16in]
V^{0-}\theta^{1}  &  =(q+q^{-2})\lambda_{+}^{-1}\theta^{1}V^{0-}%
+(q^{-1}-q^{-2})\lambda_{+}^{-1}\theta^{1}V^{3-},\\
V^{0-}\theta^{2}  &  =(q+1)\lambda_{+}^{-1}\theta^{2}V^{0-}+(q^{-1}%
-1)\lambda_{+}^{-1}\theta^{2}V^{3-}\nonumber\\
&  -\;q^{-1/2}\lambda_{+}^{-3/2}\theta^{1}\rho.\nonumber
\end{align}
For the second set of spinors we get from (\ref{VerLoGenSp})
\begin{align}
V^{+3}\bar{\theta}^{\,1}  &  =(q+q^{-1})\lambda_{+}^{-1}\bar{\theta}%
^{\,1}V^{+3}+(q-1)\lambda_{+}^{-1}\bar{\theta}^{\,1}V^{+0}\\
&  -\;q^{1/2}\lambda_{+}^{-3/2}\bar{\theta}^{\,2}\,\sigma,\nonumber\\
V^{+3}\bar{\theta}^{\,2}  &  =(q^{2}+q^{-1})\lambda_{+}^{-1}\bar{\theta}%
^{\,2}V^{+3}-(q^{2}-q)\lambda_{+}^{-1}\bar{\theta}^{\,2}V^{+0}%
,\nonumber\\[0.16in]
V^{+0}\bar{\theta}^{\,1}  &  =(q+q^{-2})\lambda_{+}^{-1}\bar{\theta}%
^{\,1}V^{+0}+(q^{-1}-q^{-2})\lambda_{+}^{-1}\bar{\theta}^{\,1}V^{+3}\\
&  -\;q^{1/2}\lambda_{+}^{-3/2}\bar{\theta}^{\,2}\,\sigma,\nonumber\\
V^{+0}\bar{\theta}^{\,2}  &  =(q+1)\lambda_{+}^{-1}\bar{\theta}^{\,2}%
V^{+0}-(1-q^{-1})\lambda_{+}^{-1}\bar{\theta}^{\,2}V^{+3},\nonumber\\[0.16in]
V^{+-}\bar{\theta}^{\,1}  &  =q^{-1}(2+q^{2}\lambda_{+})\lambda_{+}^{-2}%
\bar{\theta}^{\,1}V^{+-}-q^{-1}(q-1)^{2}\lambda_{+}^{-2}\bar{\theta}%
^{\,1}V^{30}\\
&  +\;q^{1/2}\lambda\lambda_{+}^{-3/2}\bar{\theta}^{\,2}(V^{0-}-V^{3-}%
)+\lambda_{+}^{-2}\bar{\theta}^{\,1}U^{2},\nonumber\\
V^{+-}\bar{\theta}^{\,2}  &  =q^{-1}(2+q^{2}\lambda_{+})\lambda_{+}^{-2}%
\bar{\theta}^{\,2}V^{+-}+q^{-1}(q-1)^{2}\lambda_{+}^{-2}\bar{\theta}%
^{\,2}V^{30}\nonumber\\
&  -\;q^{-1/2}\lambda(\lambda_++1)\lambda_+^{-3/2}\bar{\theta}^1V^{+0}\nonumber\\
& +\;q^{-5/2}\lambda(q^2\lambda_++1)\lambda_+^{-3/2}\bar{\theta}^1V^{+3}
-q^{-2}\lambda_{+}^{-2}\bar{\theta}^{\,2}U^{1},\nonumber\\[0.16in]
V^{30}\bar{\theta}^{\,1}  & =
q^{-1}(\lambda_++2q^2)\lambda_+^{-2}\bar{\theta}^1V^{30}
-q^{-1}(q-1)^2\lambda_+^{-2}\bar{\theta}^1V^{+-}\\
& +\;q^{3/2}\lambda\lambda_+^{-3/2}\bar{\theta}^2(V^{0-}-V^{3-})
+q\lambda_+^{-2}\bar{\theta}^1U^2,\nonumber\\
V^{30}\bar{\theta}^{\,2}  &  =q^{-1}(\lambda_++2q^2)\lambda_+^{-2}\bar{\theta}^2V^{30}
-q^{-1}(q-1)^2\lambda_+^{-2}\bar{\theta}^2V^{+-}\\
&
-\;q^{-3/2}\lambda(\lambda_++1)\lambda_+^{-3/2}\bar{\theta}^1V^{+3}\nonumber\\
&
+\;q^{-3/2}\lambda(\lambda_++q^2)\lambda_+^{-3/2}\bar{\theta}^1V^{+0}
-q^{-1}\lambda_+^{-2}\bar{\theta}^2U^2,\nonumber\\[0.16in]
V^{3-}\bar{\theta}^{\,1}  &  =(q+1)\lambda_{+}^{-1}\bar{\theta}%
^{\,1}V^{3-}+(q^2-q)\lambda_{+}^{-1}\bar{\theta}^{\,1}V^{0-},\\
V^{3-}\bar{\theta}^{\,2}  &
=(q+q^{-2})\lambda_+^{-1}\bar{\theta^2}V^{3-}
-(q-1)\lambda_+^{-1}\bar{\theta}^2V^{0-}\nonumber\\
&
+\;q^{1/2}\lambda\lambda_+^{-1/2}\bar{\theta}^1(qV^{+-}-V^{30})
+q^{-1/2}\lambda_+^{-3/2}\bar{\theta}^1\sigma,\nonumber\\[0.16in]
V^{0-}\bar{\theta}^{\,1}  &  =(q^2+q^{-1})\lambda_{+}^{-1}\bar{\theta}%
^{\,1}V^{0-}+(1-q^{-1})\lambda_{+}^{-1}\bar{\theta}^{\,1}V^{3-},\\
V^{0-}\bar{\theta}^{\,2}  &
=(1+q^{-1})\lambda_+^{-1}\bar{\theta}^2V^{0-}
-(q^{-1}-q^{-2})\lambda_+^{-1}\bar{\theta}^2V^{3-}\nonumber\\
& +\;q^{1/2}\lambda\lambda_+^{-1/2}\bar{\theta}^1(q V^{+-}-V^{30})
-q^{-5/2}\lambda_+^{-3/2}\bar{\theta}^1\sigma.\nonumber
\end{align}
The right versions of the above relations read
\begin{align}
\theta_{1}V^{+3}  &  =(q+1)\lambda_{+}^{-1}V^{+3}\theta_{1}+(q^{2}%
-q)\lambda_{+}^{-1}V^{+0}\theta_{1}\;,\\
\theta_{2}V^{+3}  &  =(q+q^{-2})\lambda_{+}^{-1}V^{+3}\theta_{2}%
-(q-1)\lambda_{+}^{-1}V^{+0}\theta_{2}\nonumber\\
&  -\;q^{1/2}\lambda\lambda_{+}^{-1/2}(V^{+-}+qV^{30})\theta_{1}%
+q^{1/2}\lambda_{+}^{-3/2}\rho\,\theta_{1}\;,\nonumber\\[0.16in]
\theta_{1}V^{+0}  &  =(q^2+q)\lambda_{+}^{-1}V^{+0}\theta_{1}%
+(1-q^{-1})\lambda_{+}^{-1}V^{+3}\theta_{1}\;,\\
\theta_{2}V^{+0}  &  =(1+q^{-1})\lambda_{+}^{-1}V^{+0}\theta_{2}%
-(q^{-1}-q^{-2})\lambda_{+}^{-1}V^{+3}\theta_{2}\nonumber\\
&  -\;q^{1/2}\lambda\lambda_{+}^{-1/2}(V^{+-}+qV^{30})\theta_{1}%
-q^{-3/2}\lambda_{+}^{-3/2}\rho\,\theta_{1}\;,\nonumber\\[0.16in]
\theta_{1}V^{+-}  &  =q^{-1}(q^{4}+1+\lambda_{+})\lambda_{+}^{-2}V^{+-}%
\theta_{1}\\
&  -\;q^{-2}(q^{4}+1-q^{2}\lambda_{+})\lambda_{+}^{-2}V^{30}\theta
_{1}\nonumber\\
&  +\;q^{-1/2}\lambda\lambda_{+}^{-3/2}(V^{+3}-V^{+0})\theta_{2}\nonumber\\
&  +\;q^{-1}\lambda\lambda_{+}^{-2}\rho\,\theta_{1}+q^{-2}\lambda_{+}%
^{-2}U^{1}\theta_{1},\;\nonumber\\
\theta_{2}V^{+-}  &  =q^{-1}(q^{4}+1+\lambda_{+})\lambda_{+}^{-2}V^{+-}%
\theta_{2}\nonumber\\
&  -\;q^{-2}(q^{4}+1-q^{2}\lambda_{+})\lambda_{+}^{-2}V^{30}\theta
_{2}\nonumber\\
&  -\;q^{-3/2}(q^{2}\lambda_{+}+1)\lambda\lambda_{+}^{-3/2}V^{3-}\theta
_{1}\nonumber\\
&  -\;q^{1/2}(\lambda_{+}-1)\lambda\lambda_{+}^{-3/2}V^{0-}\theta
_{1}\nonumber\\
&  +\;q^{-1}\lambda\lambda_{+}^{-2}\rho\,\theta_{2}-\lambda_{+}^{-2}%
U^{1}\theta_{2},\nonumber\\[0.16in]
\theta_{1}V^{30}  &  =q^{-3}(q^{4}+1+q^{4}\lambda_{+})\lambda_{+}^{-2}%
V^{30}\theta_{1}\\
&  -\;q^{-2}(q^{4}+1-q^{2}\lambda_{+})\lambda_{+}^{-2}V^{+-}\theta
_{1}\nonumber\\
&  +\;q^{-3/2}\lambda\lambda_{+}^{-3/2}(V^{+0}-V^{+3})\theta_{2}\nonumber\\
&  -\;q^{-2}\lambda\lambda_{+}^{-2}\rho\,\theta_{1}-q^{-3}\lambda_{+}%
^{-2}U^{1}\theta_{1}\;,\nonumber\\
\theta_{2}V^{30}  &  =q^{-3}(q^{4}+1+q^{4}\lambda_{+})\lambda_{+}^{-2}%
V^{30}\theta_{2}\nonumber\\
&  -\;q^{-2}(q^{4}+1-q^{2}\lambda_{+})\lambda_{+}^{-2}V^{+-}\theta
_{2}\nonumber\\
&  +\;q^{-1/2}(q^{2}\lambda_{+}+1)\lambda\lambda_{+}^{-3/2}V^{0-}\theta
_{1}\nonumber\\
&  +\;q^{-5/2}(1-q^{4}\lambda_{+})\lambda\lambda_{+}^{-3/2}V^{3-}\theta
_{1}\nonumber\\
&  -\;q^{-2}\lambda\lambda_{+}^{-2}\rho\,\theta_{2}+q^{-1}\lambda_{+}%
^{-2}U^{1}\theta_{2}\;,\nonumber\\[0.16in]
\theta_{1}V^{3-}  &  =(1+q^{-1})\lambda_{+}^{-1}V^{3-}\theta_{1}%
+(q-1)\lambda_{+}^{-1}V^{0-}\theta_{1}\\
&  -\;q^{-1/2}\lambda_{+}^{-3/2}\rho\,\theta_{2}\;,\nonumber\\
\theta_{2}V^{3-}  &  =q^{-1}(q^{3}+1)\lambda_{+}^{-1}V^{3-}\theta_{2}%
-(q^{2}-q)\lambda_{+}^{-1}V^{0-}\theta_{2}\;,\nonumber\\[0.16in]
\theta_{1}V^{0-}  &  =q^{-2}(q^{3}+1)\lambda_{+}^{-1}V^{0-}\theta_{1}%
+q^{-2}(q-1)\lambda_{+}^{-1}V^{3-}\theta_{1}\\
&  -\;q^{-1/2}\lambda_{+}^{-3/2}\rho\,\theta_{2},\;\nonumber\\
\theta_{2}V^{0-}  &  =(q+1)\lambda_{+}^{-1}V^{0-}\theta_{2}-(1-q^{-1}%
)\lambda_+^{-1}V^{3-}\theta_{2}\;,\nonumber
\end{align}
and similar for spinors with bar,\
\begin{align}
\bar{\theta}_{1}V^{+3}  &  =(1+q^{-1})\lambda_{+}^{-1}V^{+3}\bar{\theta}%
_{1}+(q-1)\lambda_{+}^{-1}V^{+0}\bar{\theta}_{1}\;,\\
\bar{\theta}_{2}V^{+3}  &  =q^{-1}(q^{3}+1)\lambda_{+}^{-1}V^{+3}\bar{\theta
}_{2}-(q^2-q)\lambda_{+}^{-1}V^{+0}\bar{\theta}_{2}\nonumber\\
&
-\;q^{1/2}\lambda_{+}^{-3/2}\sigma\,\bar{\theta}_{1}\;,\nonumber\\[0.16in]
\bar{\theta}_{1}V^{+0}  &  =q^{-2}(q^{3}+1)\lambda_{+}^{-1}V^{+0}\bar{\theta
}_{1}+(q^{-1}-q^{-2})\lambda_{+}^{-1}V^{+3}\bar{\theta}_{1}\;,\\
\bar{\theta}_{2}V^{+0}  &  =(q^{-1}-1)\lambda_{+}^{-1}V^{+3}%
\bar{\theta}_{2}+(q+1)\lambda_{+}^{-1}%
V^{+0}\bar{\theta}_{2}\nonumber\\
&
-\;q^{1/2}\lambda_{+}^{-3/2}\sigma\,\bar{\theta}_{1}\;,\nonumber\\[0.16in]
\bar{\theta}_{1}V^{+-}  &  =q^{-3}(q^{4}+1+q^{4}\lambda_{+})\lambda_{+}%
^{-2}V^{+-}\bar{\theta}_{1}\\
&  +\;q^{-1}(q-1)^{2}(\lambda_{+}+1)\lambda_{+}^{-2}V^{30}\bar{\theta}%
_{1}\nonumber\\
&  +\;(q^{2}\lambda_{+}+1)q^{-5/2}\lambda\lambda_{+}^{-3/2}V^{+3}\bar{\theta
}_{2}\nonumber\\
&  +\;q^{-1/2}(1-\lambda_{+})\lambda\lambda_{+}^{-3/2}V^{+0}\bar{\theta}%
_{2}\nonumber\\
&  -\;q^{-1}\lambda\lambda_{+}^{-2}\sigma\,\bar{\theta}_{1}+q^{-2}\lambda
_{+}^{-2}U^{2}\bar{\theta}_{1}\;,\nonumber\\
\bar{\theta}_{2}V^{+-}  &  =q^{-3}(q^{4}+1+q^{4}\lambda_{+})\lambda_{+}%
^{-2}V^{+-}\bar{\theta}_{2}\nonumber\\
&  +\;q^{-1}(q-1)^{2}(\lambda_{+}+1)\lambda_{+}^{-2}V^{30}\bar{\theta}%
_{2}\nonumber\\
&  +\;q^{1/2}\lambda\lambda_{+}^{-3/2}(V^{0-}-V^{3-})\bar{\theta}%
_{1}\nonumber\\
&  -\;q^{-1}\lambda\lambda_{+}^{-2}\sigma\,\bar{\theta}_{2}-\lambda_{+}%
^{-2}U^{2}\bar{\theta}_{2}\;,\nonumber\\[0.16in]
\bar{\theta}_{1}V^{30}  &  =q^{-1}(q^{4}+1+\lambda_{+})\lambda_{+}%
^{-2}V^{30}\bar{\theta}_{1}\\
&  +\;q^{-1}(q-1)^{2}(\lambda_{+}+1)\lambda_{+}^{-2}V^{+-}\bar{\theta}%
_{1}\nonumber\\
&  +\;q^{-3/2}(1-\lambda_{+})\lambda\lambda_{+}^{-3/2}V^{+3}\bar{\theta}%
_{2}\nonumber\\
&  +\;q^{-3/2}(q^{4}+\lambda_{+})\lambda\lambda_{+}^{-3/2}V^{+0}\bar{\theta
}_{2}\nonumber\\
&  -\;\lambda\lambda_{+}^{-2}\sigma\,\bar{\theta}_{1}+q^{-1}\lambda_{+}%
^{-2}U^{2}\bar{\theta}_{1}\;,\nonumber\\
\bar{\theta}_{2}V^{30}  &  =q^{-1}(q^{4}+1+\lambda_{+})\lambda_{+}%
^{-2}V^{30}\bar{\theta}_{2}\nonumber\\
&  +\;q^{-1}(q-1)^{2}(\lambda_{+}+1)\lambda_{+}^{-2}V^{+-}\bar{\theta}%
_{2}\nonumber\\
&  +\;q^{3/2}\lambda\lambda_{+}^{-3/2}(V^{0-}-V^{3-})\bar{\theta}%
_{1}\nonumber\\
&  -\;\lambda\lambda_{+}^{-2}\sigma\,\bar{\theta}_{2}-q\lambda_{+}^{-2}%
U^{2}\bar{\theta}_{2}\;,\nonumber\\[0.16in]
\bar{\theta}_{1}V^{3-}  &  =(q+1)\lambda_{+}^{-1}V^{3-}\bar{\theta}_{1}%
+(q^{2}-q)\lambda_{+}^{-1}V^{0-}\bar{\theta}_{1}\\
&  +\;q^{1/2}\lambda\lambda_{+}^{-1/2}(qV^{+-}-V^{30})\bar{\theta}%
_{2}+q^{-1/2}\lambda_{+}^{-3/2}\sigma\,\bar{\theta}_{2}\;,\nonumber\\
\bar{\theta}_{2}V^{3-}  &  =q^{-2}(q^{3}+1)\lambda_{+}^{-1}V^{3-}\bar{\theta
}_{2}-(q-1)\lambda_{+}^{-1}V^{0-}\bar{\theta}_{2}\;,\nonumber\\[0.16in]
\bar{\theta}_{1}V^{0-}  &  =q^{-1}(q^{3}+1)\lambda_{+}^{-1}V^{0-}\bar{\theta
}_{1}+(1-q^{-1})\lambda_{+}^{-1}V^{3-}\bar{\theta}_{1}\\
&  +\;q^{1/2}\lambda\lambda_{+}^{-1/2}(qV^{+-}-V^{30})\bar{\theta}%
_{2}-q^{-5/2}\lambda_{+}^{-3/2}\sigma\,\bar{\theta}_{2}\;,\nonumber\\
\bar{\theta}_{2}V^{0-}  &  =(q+q^{-1})\lambda_{+}^{-1}V^{0-}\bar{\theta}%
_{2}-(q^{-1}-q^{-2})\lambda_{+}^{-1}V^{3-}\bar{\theta}_{2}\;.\nonumber
\end{align}

Next, we turn to the vector representations for the $V^{\mu\nu},$ which are
seen to be
\begin{equation}
V^{\mu\nu}\rhd X^{\rho}=(\tau^{\mu\nu})^{\rho}{}_{\sigma}\;X^{\sigma},\quad
X_{\rho}\lhd V^{\mu\nu}=X_{\sigma}(\tau^{\mu\nu})^{\sigma}{}_{\rho},
\end{equation}
with
\begin{align}
(\tau^{+3})^{\rho}{}_{\sigma}  &  =\lambda_{+}^{-2}\left(
\begin{array}
[c]{cccc}%
0 & -2q & -\lambda & 0\\
0 & 0 & 0 & -2\\
0 & 0 & 0 & -q\lambda\\
0 & 0 & 0 & 0
\end{array}
\right)  ,\\
(\tau^{+0})^{\rho}{}_{\sigma}  &  =\lambda_{+}^{-2}\left(
\begin{array}
[c]{cccc}%
0 & -\lambda & 2q^{-1} & 0\\
0 & 0 & 0 & -q^{-1}\lambda\\
0 & 0 & 0 & -(q^{2}+q^{-2})\\
0 & 0 & 0 & 0
\end{array}
\right)  ,\nonumber\\
(\tau^{+-})^{\rho}{}_{\sigma}  &  =\lambda_{+}^{-2}\left(
\begin{array}
[c]{cccc}%
2q^{-2} & 0 & 0 & 0\\
0 & -2q^{-1}\lambda & -q^{-1}\lambda & 0\\
0 & q^{-1}\lambda & 0 & 0\\
0 & 0 & 0 & -2
\end{array}
\right)  ,\nonumber\\
(\tau^{30})^{\rho}{}_{\sigma}  &  =\lambda_{+}^{-2}\left(
\begin{array}
[c]{cccc}%
\lambda q^{-2} & 0 & 0 & 0\\
0 & -q^{-1}\lambda^{2} & 2q & 0\\
0 & q^{-1}(q^{2}+q^{-2}) & 0 & 0\\
0 & 0 & 0 & -\lambda
\end{array}
\right)  ,\nonumber\\
(\tau^{3-})^{\rho}{}_{\sigma}  &  =\lambda_{+}^{-2}\left(
\begin{array}
[c]{cccc}%
0 & 0 & 0 & 0\\
2 & 0 & 0 & 0\\
-q^{-1}\lambda & 0 & 0 & 0\\
0 & 2q^{-1} & -\lambda & 0
\end{array}
\right)  ,\nonumber\\
(\tau^{0-})^{\rho}{}_{\sigma}  &  =\lambda_{+}^{-2}\left(
\begin{array}
[c]{cccc}%
0 & 0 & 0 & 0\\
q^{-1}\lambda & 0 & 0 & 0\\
2q^{-2} & 0 & 0 & 0\\
0 & q^{-2}\lambda & -q^{-1}(q^{2}+q^{-2}) & 0
\end{array}
\right)  .\nonumber
\end{align}
Notice that rows and columns are labeled in the order $+,3,0,-$. For $U^{1}$,
$U^{2}$, $\sigma$ and $\rho$ we obtain the matrices%
\begin{align}
{(U^{1})^{\rho}}_{\sigma}  &  ={(U^{1})^{\rho}}_{\sigma}=-(q^{2}%
+q^{-2})\lambda_{+}^{-1}{\delta^{\rho}}_{\sigma},\\
{(\rho)^{\rho}}_{\sigma}  &  =\lambda_{+}^{-1}%
\begin{pmatrix}
-q\lambda_{+} & 0 & 0 & 0\\
0 & -2 & q\lambda & 0\\
0 & q^{-1}\lambda & -(q^{2}+q^{-2}) & 0\\
0 & 0 & 0 & -q^{-1}\lambda_{+}%
\end{pmatrix}
,\nonumber\\
{(\sigma)^{\rho}}_{\sigma}  &  =\lambda_{+}^{-1}%
\begin{pmatrix}
q\lambda_{+} & 0 & 0 & 0\\
0 & 2 & q^{-1}\lambda & 0\\
0 & q\lambda & (q^{2}+q^{-2}) & 0\\
0 & 0 & 0 & q^{-1}\lambda_{+}%
\end{pmatrix}
,\nonumber
\end{align}
and likewise for the generators of the $U_{q}(su_{2})$-subalgebra,%
\begin{gather}
{(L}^{+}{)^{\rho}}_{\sigma}=%
\begin{pmatrix}
0 & -q & 0 & 0\\
0 & 0 & 0 & -1\\
0 & 0 & 0 & 0\\
0 & 0 & 0 & 0
\end{pmatrix}
,\quad{(L}^{-}{)^{\rho}}_{\sigma}=%
\begin{pmatrix}
0 & 0 & 0 & 0\\
1 & 0 & 0 & 0\\
0 & 0 & 0 & 0\\
0 & q^{-1} & 0 & 0
\end{pmatrix}
,\\
{(\tau}^{-1/2}{)^{\rho}}_{\sigma}=%
\begin{pmatrix}
q^{2} & 0 & 0 & 0\\
0 & 1 & 0 & 0\\
0 & 0 & 1 & 0\\
0 & 0 & 0 & q^{-2}%
\end{pmatrix}
.\nonumber
\end{gather}

We can now proceed to write down commutation relations between the $V^{\mu\nu
}$ and the components of a vector operator. Let us recall that we mean by a
vector operator a set of objects $X^{\rho}$ which transform under Lorentz
transformations according to\
\begin{equation}
{[}V^{\mu\nu},X^{\rho}{]}_{q}=(\tau^{\mu\nu})^{\rho}{}_{\sigma}\,X^{\sigma
},\quad{[}X_{\sigma},V^{\mu\nu}{]}_{q}=X_{\sigma}\,(\tau^{\mu\nu})^{\sigma}%
{}_{\rho}.
\end{equation}
These relations are equivalent to those in (\ref{VerLoGenSp}) if we substitute
the vector components for the general element $a$. By taking the coproduct of
the ${V}^{\mu\nu}$ together with the vector representations of the Lorentz
generators we can again compute from (\ref{VerLoGenSp}) the explicit form for
the commutation relations between the ${V}^{\mu\nu}$ and the $X^{\rho}.$
Unfortunately, the results are rather lengthy:
\begin{align}
V^{+3}X^{+}  &  =q^{-2}(q^{4}+1)\lambda_{+}^{-1}X^{+}V^{+3}-q\lambda
\lambda_{+}^{-1}X^{+}V^{+0},\\
V^{+3}X^{3}  &  =2\lambda_{+}^{-1}X^{3}V^{+3}-q\lambda X^{+}V^{+-}%
+q\lambda\lambda_{+}^{-1}X^{0}V^{+3}\nonumber\\
&  -q\lambda\lambda_{+}^{-1}X^{+}V^{30}+q\lambda_{+}^{-2}X^{+}(U^{1}%
+U^{2}),\nonumber\\
V^{+3}X^{0}  &  =q^{-2}(q^{4}+1)\lambda_{+}^{-1}X^{0}V^{+3}+q^{-1}\lambda
\lambda_{+}^{-1}X^{3}V^{+3}\nonumber\\
&  -q^{2}\lambda\lambda_{+}^{-1}X^{+}V^{+-}+\lambda_{+}^{-2}X^{+}%
(qU^{1}-q^{-1}U^{2}),\nonumber\\
V^{+3}X^{-}  &  =2\lambda_{+}^{-1}X^{-}V^{+3}-2q\lambda\lambda_{+}^{-1}%
X^{3}V^{+-}+q\lambda\lambda_{+}^{-1}X^{0}V^{+-}\nonumber\\
&  +\;q\lambda\lambda_{+}^{-1}X^{-}V^{+0}-q^{2}\lambda\lambda_{+}^{-1}%
X^{3}V^{30}\nonumber\\
& -\;q^{2}\lambda^{2}\lambda_{+}^{-1}X^{+}V^{0-}+q^2\lambda^2\lambda_+^{-1}X^+V^{3-}\nonumber\\
&  +\lambda_{+}^{-2}X^{3}(U^{1}+U^{2})-\lambda_{+}^{-2}X^{0}(U^{1}-q^{2}%
U^{2}),\nonumber\\[0.2in]
V^{+0}X^{+}  &  =2\lambda_{+}^{-1}X^{+}V^{+0}-q^{-1}\lambda\lambda_{+}^{-1}%
X^{+}V^{+3},\\
V^{+0}X^{3}  &  =2\lambda_{+}^{-1}X^{3}V^{+0}-\lambda\lambda_{+}^{-1}%
X^{+}V^{+3}-2q\lambda\lambda_{+}^{-1}X^{+}V^{30}\nonumber\\
&  +q\lambda\lambda_{+}^{-1}X^{0}V^{+0}-\lambda_{+}^{-2}X^{+}(q^{-1}U^{1}%
-qU^{2}),\nonumber\\
V^{+0}X^{0}  &  =q^{-2}\lambda_{+}^{-1}(q^{4}+1)X^{0}V^{+0}-q\lambda
\lambda_{+}^{-1}X^{+}V^{30}\nonumber\\
&  +q^{-1}\lambda\lambda_{+}^{-1}X^{3}V^{+0}-q^{-1}\lambda_{+}^{-2}X^{+}%
(U^{1}+U^{2}),\nonumber\\
V^{+0}X^{-}  &  =q^{-2}\lambda_{+}^{-1}(q^{4}+1)X^{-}V^{+0}-q\lambda
X^{3}V^{30}+\lambda\lambda_{+}^{-1}X^{0}V^{30}\nonumber\\
&  +q^{-1}\lambda\lambda_{+}^{-1}X^{-}V^{+3}-q\lambda\lambda_{+}^{-1}%
X^{3}V^{+-}-q^{2}\lambda^{2}\lambda_{+}^{-1}X^{+}V^{0-}\nonumber\\
&  +q^{2}\lambda^{2}\lambda_{+}^{-1}X^{+}V^{3-}+\lambda_{+}^{-2}X^{0}%
(q^{-2}U^{1}+q^{2}U^{2})\nonumber\\
&  -\lambda_{+}^{-2}X^{3}(q^{-2}U^{1}-U^{2}),\nonumber\\[0.2in]
V^{+-}X^{+}  &  =2\lambda_{+}^{-1}X^{+}V^{+-}+q^{-1}\lambda X^{3}%
V^{+3}-\lambda\lambda_{+}^{-1}X^{3}V^{+0}\\
&  -q^{-2}\lambda\lambda_{+}^{-1}X^{0}V^{+3}-q^{-2}\lambda_{+}^{-2}X^{+}%
(U^{1}+U^{2}),\nonumber\\
V^{+-}X^{3}  &  =(2-\lambda^{2})\lambda_{+}^{-1}X^{3}V^{+-}-\lambda
X^{+}V^{3-}+q\lambda\lambda_{+}^{-1}X^{+}V^{0-}\nonumber\\
&  +2q^{-1}\lambda\lambda_{+}^{-1}X^{-}V^{+3}-q^{-1}\lambda\lambda_{+}%
^{-1}X^{-}V^{+0}+\lambda^{2}\lambda_{+}^{-1}X^{0}V^{+-}\nonumber\\
&  +q^{-2}\lambda_{+}^{-2}X^{0}U^{1}%
-U^{2})+q^{-1}\lambda\lambda_{+}^{-2}X^{3}(U^{1}%
+U^{2}),\nonumber\\
V^{+-}X^{0}  &  =q^{-2}(q^{4}+1)\lambda_{+}^{-1}X^{0}V^{+-}-\lambda^{2}%
\lambda_{+}^{-1}X^{3}V^{+-}+q^{-1}\lambda\lambda_{+}^{-1}X^{-}V^{+3}%
\nonumber\\
&  -q^{-1}\lambda\lambda_{+}^{-1}X^{+}V^{3-}+\;\lambda_{+}^{-2}X^{3}%
(U^{1}-q^{-2}U^{2}),\nonumber\\
V^{+-}X^{-}  &  =2\lambda_{+}^{-1}X^{-}V^{+-}-2\lambda\lambda_{+}^{-1}%
X^{3}V^{3-}+\lambda\lambda_{+}^{-1}X^{3}V^{0-}\nonumber\\
&  +\lambda\lambda_{+}^{-1}X^{0}V^{3-}+\lambda_{+}^{-2}X^{-}(U^{1}%
+U^{2}),\nonumber\\[0.2in]
V^{30}X^{+}  &  =2\lambda_{+}^{-1}X^{+}V^{30}-q^{-1}\lambda\lambda_{+}%
^{-1}X^{0}V^{+0}-q^{-1}\lambda\lambda_{+}^{-1}X^{3}V^{+3}\\
&  +\;2q^{-1}\lambda\lambda_{+}^{-1}X^{3}V^{+0}+q^{-1}\lambda_{+}^{-2}%
X^{+}(q^{-2}U^{1}-U^{2}),\nonumber\\
V^{30}X^{3}  &  =(2-\lambda^{2})\lambda_{+}^{-1}X^{3}V^{30}+q\lambda
X^{+}V^{0-}+q^{-1}\lambda X^{-}V^{+0}\nonumber\\
&  +\;q^{-2}\lambda\lambda_{+}^{-1}(1-q^{3}\lambda_{+})X^{+}V^{3-}%
-q^{-2}\lambda\lambda_{+}^{-1}X^{-}V^{+3}\nonumber\\
&  +\lambda^{2}\lambda_{+}^{-1}X^{0}V^{30}-\lambda\lambda_{+}^{-2}X^{3}%
(q^{-2}U^{1}-U^{2})\nonumber\\
&  -\lambda_{+}^{-2}X^{0}(q^{-3}U^{1}+qU^{2}),\nonumber\\
V^{30}X^{0}  &  =q^{-2}(q^{4}+1)\lambda_{+}^{-1}X^{0}V^{30}+\lambda\lambda
_{+}^{-1}X^{+}V^{0-}+\lambda\lambda_{+}^{-1}X^{-}V^{+0}\nonumber\\
&  -\;q^{-1}\lambda^{2}\lambda_{+}^{-1}X^{+}V^{3-}-\lambda^{2}\lambda_{+}%
^{-1}X^{3}V^{30}-q^{-1}\lambda_{+}^{-2}X^{3}(U^{1}+U^{2}),\nonumber\\
V^{30}X^{-}  &  =2\lambda_{+}^{-1}X^{-}V^{30}-\lambda\lambda_{+}%
^{-1}(q+1)X^{3}V^{3-}-q\lambda\lambda_{+}^{-1}X^{0}V^{0-}\nonumber\\
&  +2q\lambda\lambda_{+}^{-1}X^{3}V^{0-}+\lambda^{2}\lambda_{+}^{-1}%
X^{0}V^{3-}-\lambda_{+}^{-2}X^{-}(q^{-1}U^{1}-qU^{2}),\nonumber\\[0.2in]
V^{3-}X^{+}  &  =q^{-2}(q^{4}+1)\lambda_{+}^{-1}X^{+}V^{3-}-q\lambda
\lambda_{+}^{-1}X^{0}V^{+-}\\
&  -\;q\lambda\lambda_{+}^{-1}X^{+}V^{0-}+2q\lambda\lambda_{+}^{-1}X^{3}%
V^{+-}+\lambda^{2}\lambda_{+}^{-1}X^{-}(V^{+3}-V^{+0})\nonumber\\
&  -\lambda\lambda_+^{-1}X^3V^{30}-\lambda_{+}^{-2}X^{3}(U^{1}+U^{2})-\lambda_{+}^{-2}X^{0}(q^{-2}U^{1}%
-U^{2}),\nonumber\\
V^{3-}X^{3}  &  =2\lambda_{+}^{-1}X^{3}V^{3-}+q\lambda X^{-}V^{+-}%
-q^{-1}\lambda\lambda_{+}^{-1}X^{0}V^{3-}\nonumber\\
&  -q\lambda\lambda_{+}^{-1}X^{-}V^{30}-q^{-1}\lambda_{+}^{-2}X^{-}%
(U^{1}+U^{2}),\nonumber\\
V^{3-}X^{0}  &  =q^{-2}(q^{4}+1)\lambda_{+}^{-1}X^{0}V^{3-}-q\lambda
\lambda_{+}^{-1}X^{3}V^{3-}\nonumber\\
&  +\lambda\lambda_{+}^{-1}X^{-}V^{+-}+\lambda_{+}^{-2}X^{-}(qU^{1}%
-q^{-1}U^{2}),\nonumber\\
V^{3-}X^{-}  &  =2\lambda_{+}^{-1}X^{-}V^{3-}+q\lambda\lambda_{+}^{-1}%
X^{-}V^{0-},\nonumber\\[0.2in]
V^{0-}X^{+}  &  =2\lambda_{+}^{-1}X^{+}V^{0-}+\lambda\lambda_{+}%
^{-1}(q+1)X^{3}V^{+-}-2\lambda\lambda_{+}^{-1}X^{3}V^{30}\\
&  -q^{-1}\lambda\lambda_{+}^{-1}X^{+}V^{3-}+\lambda\lambda_{+}^{-1}%
X^{0}V^{30}\nonumber\\
&  +\lambda^{2}\lambda_{+}^{-1}X^{-}(V^{+3}-V^{+0})-\lambda^{2}\lambda
_{+}^{-1}X^{0}V^{+-}\nonumber\\
&  -\lambda_{+}^{-2}X^{3}(U^{1}-q^{-2}U^{2})-q^{-2}\lambda_{+}^{-2}X^{0}%
(U^{1}+U^{2}),\nonumber\\
V^{0-}X^{3}  &  =2\lambda_{+}^{-1}X^{3}V^{0-}-\lambda X^{-}V^{30}%
-q^{-2}\lambda\lambda_{+}^{-1}X^{-}(1-q^3\lambda_{+})V^{+-}\nonumber\\
&  -q^{-1}\lambda\lambda_{+}^{-1}X^{0}V^{0-}-q^{-1}\lambda_{+}^{-2}X^{-}%
(U^{1}-q^{-2}U^{2}),\nonumber\\
V^{0-}X^{0}  &  =q^{-2}(q^{4}+1)\lambda_{+}^{-1}X^{0}V^{0-}-q\lambda
\lambda_{+}^{-1}X^{3}V^{0-}\nonumber\\
&  -q^{-1}\lambda\lambda_{+}^{-1}X^{-}V^{30}+q^{-1}\lambda^{2}\lambda_{+}%
^{-1}X^{-}V^{+-}\nonumber\\
&  +\lambda_{+}^{-2}X^{-}(qU^{1}+q^{-3}U^{2}),\nonumber\\
V^{0-}X^{-}  &  =q^{-2}(q^{4}+1)\lambda_{+}^{-1}X^{-}V^{0-}+q^{-1}%
\lambda\lambda_{+}^{-1}X^{-}V^{3-}.\nonumber
\end{align}
If we want to commute the $V^{\mu\nu}$ to the left of a vector operator, we
have to apply instead
\begin{align}
X_{+}V^{+3}  &  =q^{-2}(q^{4}+1)\lambda_{+}^{-1}V^{+3}X_{+}-q\lambda
V^{+-}X_{3}\\
&  -q\lambda\lambda_{+}^{-1}V^{30}X_{3}-q\lambda\lambda_{+}^{-1}V^{+0}%
X_{+}-q^{2}\lambda\lambda_{+}^{-1}V^{+-}X_{0}\nonumber\\
&  +q^{2}\lambda^{2}\lambda_{+}^{-1}(V^{3-}-V^{0-})X_{-}+q\lambda_{+}%
^{-2}(U^{1}+U^{2})X_{3}\nonumber\\
&  +\lambda_{+}^{-2}(qU^{1}-q^{-1}U^{2})X_{0}\;,\nonumber\\
X_{3}V^{+3}  &  =2\lambda_{+}^{-1}V^{+3}X_{3}-2q\lambda\lambda_{+}^{-1}%
V^{+-}X_{-}+q^{-1}\lambda\lambda_{+}^{-1}V^{+3}X_{0}\nonumber\\
&  -q^{2}\lambda\lambda_{+}^{-1}V^{30}X_{-}+\lambda_{+}^{-2}(U^{1}+U^{2}%
)X_{-}\;,\nonumber\\
X_{0}V^{+3}  &  =q^{-2}(q^{4}+1)\lambda_{+}^{-1}V^{+3}X_{0}+q\lambda
\lambda_{+}^{-1}V^{+3}X_{3}\nonumber\\
&  +q\lambda\lambda_{+}^{-1}V^{+-}X_{-}-\lambda_{+}^{-2}(U^{1}-q^{2}%
U^{2})X_{-}\;,\nonumber\\
X_{-}V^{+3}  &  =2\lambda_{+}^{-1}V^{+3}X_{-}+q\lambda\lambda_{+}^{-1}%
V^{+0}X_{-}\;,\nonumber\\[0.2in]
X_{+}V^{+0}  &  =2\lambda_{+}^{-1}V^{+0}X_{+}-\lambda\lambda_{+}^{-1}%
V^{+-}X_{3}-q^{-1}\lambda\lambda_{+}^{-1}V^{+3}X_{+}\\
&  -2q\lambda\lambda_{+}^{-1}V^{30}X_{3}-q\lambda\lambda_{+}^{-1}V^{30}%
X_{0}+q^{2}\lambda^{2}\lambda_{+}^{-2}(V^{3-}-V^{0-})X_{-}\nonumber\\
&  -\lambda_{+}^{-2}(q^{-1}U^{1}-qU^{2})X_{3}-q^{-1}\lambda_{+}^{-2}%
(U^{1}+U^{2})X_{0}\;,\nonumber\\
X_{3}V^{+0}  &  =2\lambda_{+}^{-1}V^{+0}X_{3}-q\lambda V^{30}X_{-}%
+q^{-1}\lambda\lambda_{+}^{-1}V^{+0}X_{0}\nonumber\\
&  -q\lambda\lambda_{+}^{-1}V^{+-}X_{-}-\lambda_{+}^{-2}(q^{-2}U^{1}%
-U^{2})X_{-}\;,\nonumber\\
X_{0}V^{+0}  &  =q^{-2}(q^{4}+1)\lambda_{+}^{-1}V^{+0}X_{0}+\lambda\lambda
_{+}^{-1}V^{30}X_{-}\nonumber\\
&  +q\lambda\lambda_{+}^{-1}V^{+0}X_{3}+\lambda_{+}^{-2}(q^{-2}U^{1}%
+q^{2}U^{2})X_{-}\;,\nonumber\\
X_{-}V^{+0}  &  =q^{-2}(q^{4}+1)\lambda_{+}^{-1}%
V^{+0}X_-+q^{-1}\lambda\lambda_{+}^{-1}V^{+3}X_{-}\;,\nonumber\\[0.2in]
X_{+}V^{+-}  &  =2\lambda_{+}^{-1}V^{+-}X_{+}-\lambda V^{3-}X_{3}%
-q^{-1}\lambda\lambda_{+}^{-1}V^{3-}X_{0}\\
&  +q\lambda\lambda_{+}^{-1}V^{0-}X_{3}-q^{-2}\lambda_{+}^{-2}(U^{1}%
+U^{2})X_{+}\;,\nonumber\\
X_{3}V^{+-}  &  =(2-\lambda^{2})\lambda_{+}^{-1}V^{+-}X_{3}+q^{-1}\lambda
V^{+3}X_{+}-\lambda\lambda_{+}^{-1}V^{+0}X_{+}\nonumber\\
&  +\lambda\lambda_{+}^{-1}V^{0-}X_{-}-2\lambda\lambda_{+}^{-1}V^{3-}%
X_{-}-\lambda^{2}\lambda_{+}^{-1}V^{+-}X_{0}\nonumber\\
&  +q^{-1}\lambda\lambda_{+}^{-1}(U^{1}+U^{2})X_{3}+\lambda_{+}^{-2}%
(U^{1}-q^{-2}U^{2})X_{0},\nonumber\\
X_{0}V^{+-}  &  =q^{-2}(q^{4}+1)\lambda_{+}^{-1}V^{+-}X_{0}-q^{-2}%
\lambda\lambda_{+}^{-1}V^{+3}X_{+}\nonumber\\
&  +\lambda\lambda_{+}^{-1}V^{3-}X_{-}+\lambda^{2}\lambda_{+}^{-1}V^{+-}%
X_{3}+\lambda_{+}^{-2}(q^{-2}U^{1}-U^{2})X_{3}\;,\nonumber\\
X_{-}V^{+-}  &  =2\lambda_{+}^{-1}V^{+-}X_{-}+2q^{-1}\lambda\lambda_{+}%
^{-1}V^{+3}X_{3}+q^{-1}\lambda\lambda_{+}^{-1}V^{+3}X_{0}\nonumber\\
&  -q^{-1}\lambda\lambda_{+}^{-1}V^{+0}X_{3}+\lambda_{+}^{-2}(U^{1}%
+U^{2})X_{-}\;,\nonumber\\[0.2in]
X_{+}V^{30}  &  =2\lambda_{+}^{-1}V^{30}X_{+}+q^{-2}\lambda\lambda_{+}%
^{-1}(1-q^{3}\lambda_{+})V^{3-}X_{3}\\
&  +q\lambda V^{0-}X_{3}+\lambda\lambda_{+}^{-1}V^{0-}X_{0}-q^{-1}\lambda
^{2}\lambda_{+}^{-1}V^{3-}X_{0}\nonumber\\
&  +q^{-1}\lambda_{+}^{-2}(q^{-2}U^{1}-U^{2})X_{+}\;,\nonumber\\
X_{3}V^{30}  &  =(2-\lambda^{2})\lambda_{+}^{-1}V^{30}X_{3}-(q+\lambda
)\lambda\lambda_{+}^{-1}V^{3-}X_{-}\nonumber\\
&  -q^{-1}\lambda\lambda_{+}^{-1}V^{+3}X_{+}+2\lambda\lambda_{+}^{-1}%
(q^{-1}V^{+0}+qV^{0-})X_{+}\nonumber\\
&  -\lambda^{2}\lambda_{+}V^{30}X_{0}-\lambda\lambda_{+}^{-1}(q^{-2}%
U^{1}-U^{2})X_{3}\nonumber\\
&  +q^{-1}\lambda_{+}^{-2}(U^{1}-U^{2})X_{0}\;,\nonumber\\
X_{0}V^{30}  &  =q^{-2}(q^{4}+1)V^{30}X_{0}-q^{-1}\lambda\lambda_{+}%
^{-1}V^{+0}X_{+}-q\lambda\lambda_{+}^{-1}V^{0-}X_{-}\nonumber\\
&  +\lambda^{2}\lambda_{+}^{-1}V^{30}X_{3}+\lambda^{2}\lambda_{+}^{-1}%
V^{3-}X_{-}-\lambda_{+}^{-2}(q^{-3}U^{1}+qU^{2})X_{3}\;,\nonumber\\
X_{-}V^{30}  &  =2\lambda_{+}^{-1}V^{30}X_-+q^{-1}\lambda
V^{+0}X_{0}-q^{-2}\lambda\lambda_{+}^{-1}V^{+3}X_{3}\nonumber\\
&  +\lambda\lambda_{+}^{-1}V^{+0}X_{0}-\lambda_{+}^{-2}(q^{-1}U^{1}%
-qU^{2})X_{-}\;,\nonumber\\[0.2in]
X_{+}V^{3-}  &  =q^{-2}(q^{4}+1)V^{3-}X_{+}-q\lambda\lambda_{+}^{-1}%
V^{0-}X_{+}\;,\\
X_{3}V^{3-}  &  =2\lambda_{+}^{-1}V^{3-}X_{3}+2q\lambda\lambda_{+}^{-1}%
V^{+-}X_{+}-\lambda\lambda_{+}^{-1}V^{30}X_{+}\nonumber\\
&  -q\lambda\lambda_{+}^{-1}V^{3-}X_{0}-\lambda_{+}^{-2}(U^{1}+U^{2}%
)X_{+}\;,\nonumber\\
X_{0}V^{3-}  &  =q^{-2}(q^{4}+1)\lambda_{+}^{-1}V^{3-}X_{0}-q\lambda
\lambda_{+}^{-1}V^{+-}X_{+}\nonumber\\
&  -q^{1}\lambda\lambda_{+}^{-1}V^{3-}X_{3}-\lambda_{+}^{-2}(q^{-2}U^{1}%
-U^{2})X_{+}\;,\nonumber\\
X_{-}V^{3-}  &  =2\lambda_{+}^{-1}V^{3-}X_{-}+q\lambda V^{+-}X_{3}%
+\lambda\lambda_{+}^{-1}V^{+-}X_{0}\nonumber\\
&  +\lambda^{2}\lambda_{+}^{-1}(V^{+3}-V^{+0})X_{+}+q\lambda\lambda_{+}%
^{-1}V^{0-}X_{-}-q\lambda\lambda_{+}^{-1}V^{30}X_{3}\nonumber\\
&  +\lambda_{+}^{-2}(qU^{1}-q^{-1}U^{2})X_{0}-q^{-1}\lambda_{+}^{-2}%
(U^{1}+U^{2})X_{3}\;,\nonumber\\[0.2in]
X_{+}V^{0-}  &  =2\lambda_{+}^{-1}V^{0-}X_{+}-q^{-1}\lambda\lambda_{+}%
^{-1}V^{3-}X_{+}\;,\\
X_{3}V^{0-}  &  =2\lambda_{+}^{-1}V^{0-}X_{3}+\lambda(q+\lambda)\lambda
_{+}^{-1}V^{+-}X_{+}-2\lambda\lambda_{+}^{-1}V^{30}X_{+}\nonumber\\
&  -q\lambda\lambda_{+}^{-1}V^{0-}X_{0}-\lambda_{+}^{-2}(U^{1}-q^{-2}%
U^{2})X_{+}\;,\nonumber\\
X_{0}V^{0-}  &  =q^{-2}(q^{4}+1)\lambda_{+}^{-1}V^{0-}X_{0}+\lambda\lambda
_{+}^{-1}V^{30}X_{+}\nonumber\\
&  -q^{-1}\lambda\lambda_{+}^{-1}V^{0-}X_{3}-\lambda^{2}\lambda_{+}^{-1}%
V^{+-}X_{+}-q^{-2}\lambda_{+}^{-2}(U^{1}+U^{2})X_{+}\;,\nonumber\\
X_{-}V^{0-}  &  =q^{-2}(q^{4}+1)\lambda_{+}^{-1}V^{0-}X_{-}-\lambda
V^{30}X_{3}+q^{-1}\lambda\lambda_{+}^{-1}V^{3-}X_{-}\nonumber\\
&  -q^{-1}\lambda\lambda_{+}^{-1}V^{30}X_{0}+q^{-3}\lambda\lambda_{+}%
^{-1}(-q+q^{-2}\lambda_{+})V^{+-}X_{3}\nonumber\\
&  -q^{-1}\lambda_{+}^{-2}(U^{1}-q^{-1}U^{2})X_{3}+\lambda_{+}^{-2}%
(qU^{1}+q^{-2}U^{2})X_{0}.\nonumber
\end{align}

\subsection{Quantum Lie algebra of q-deformed Lorentz algebra and its Casimir
operators}

Last but not least we would like to present the quantum Lie algebra of
q-deformed Lorentz algebra. For this to achieve, we calculate the adjoint
action of the independent $V^{\mu\nu}$ on themselves in a fashion as was done
in the previous section and set the results equal to the q-commutators of the
corresponding Lorentz generators. In the cases where the adjoint actions do
not vanish, we obtain
\begin{align}
{[}V^{+3},V^{+-}{]}_{q}  &  =-q^{2}{[}V^{+-},V^{+3}{]}_{q}=-\lambda_{+}%
^{-1}V^{+3},\\
{[}V^{+3},V^{30}{]}_{q}  &  =-q^{2}{[}V^{30},V^{+3}{]}_{q}=-q\lambda_{+}%
^{-1}V^{+0},\nonumber\\
{[}V^{+3},V^{3-}{]}_{q}  &  =-{[}V^{3-},V^{+3}{]}_{q}=-q\lambda_{+}^{-1}%
V^{+-},\nonumber\\
{[}V^{+3},V^{0-}{]}_{q}  &  =-{[}V^{0-},V^{+3}{]}_{q}=\lambda_{+}^{-1}%
(V^{30}-\lambda V^{+-}),\nonumber\\[0.16in]
{[}V^{+0},V^{+-}{]}_{q}  &  =-q^{2}{[}V^{+-},V^{+0}{]}_{q}=-\lambda_{+}%
^{-1}V^{+0},\\
{[}V^{+0},V^{30}{]}_{q}  &  =-q^{2}{[}V^{30},V^{+0}{]}_{q}=-\lambda_{+}%
^{-1}(q^{-1}V^{+3}+\lambda V^{+0}),\nonumber\\
{[}V^{+0},V^{3-}{]}_{q}  &  =-{[}V^{3-},V^{+0}{]}_{q}=-\lambda_{+}^{-1}%
V^{30},\nonumber\\
{[}V^{+0},V^{0-}{]}_{q}  &  =-{[}V^{0-},V^{+-}{]}_{q}=q^{-1}\lambda_{+}%
^{-1}V^{+-},\nonumber\\[0.16in]
{[}V^{+-},V^{+-}{]}_{q}  &  =-q^{-1}\lambda\lambda_{+}^{-1}V^{+-},\\
{[}V^{+-},V^{30}{]}_{q}  &  ={[}V^{30},V^{+-}{]}_{q}=-q^{-1}\lambda\lambda
_{+}^{-1}V^{30},\nonumber\\
{[}V^{+-},V^{3-}{]}_{q}  &  =-q^{2}{[}V^{3-},V^{+-}{]}_{q}=-\lambda_{+}%
^{-1}V^{3-},\nonumber\\
{[}V^{+-},V^{0-}{]}_{q}  &  =-q^{2}{[}V^{0-},V^{+-}{]}_{q}=-\lambda_{+}%
^{-1}V^{0-},\nonumber\\[0.16in]
{[}V^{30},V^{30}{]}_{q}  &  =-q^{-1}\lambda\lambda_{+}^{-1}(V^{+-}+\lambda
V^{30}),\\
{[}V^{30},V^{3-}{]}_{q}  &  =-q^{2}{[}V^{3-},V^{30}{]}_{q}=\lambda_{+}^{-1}(qV^{0-}-\lambda V^{3-}%
),\nonumber\\
{[}V^{30},V^{0-}{]}_{q}  &  =-q^{2}{[}V^{0-},V^{30}{]}_{q}=q^{-1}\lambda
_{+}^{-1}V^{3-}.\nonumber
\end{align}
Of course, the quantum Lie algebra relations are consistent with spinor and
vector representation of q-deformed Lorentz algebra. This can again be checked
in a familiar\ way by writing out q-commutators and substituting the
representation matrices for the Lorentz generators.

As Casimirs of this quantum Lie algebra we have found the two operators
\begin{align}
C^{1}  &  =\eta_{\mu\nu}\eta_{\rho\sigma}V^{\mu\rho}V^{\nu\sigma}%
=2V^{30}V^{30}+(q^{2}+q^{-2})V^{+-}V^{+-}\\
&  +\;2(qV^{+0}V^{0-}-q^{-1}V^{+3}V^{3-}-q^{-3}V^{3-}V^{3+})\nonumber\\
&  -\;\lambda(V^{+3}V^{0-}+V^{+0}V^{3-}+V^{+-}V^{30}+V^{30}V^{+-})\nonumber\\
&  -\;q^{-2}\lambda(V^{3-}V^{+0}+V^{0-}V^{+3}),\nonumber\\[0.16in]
C^{2}  &  =\varepsilon_{\mu\nu\rho\sigma}V^{\mu\nu}V^{\rho\sigma}%
=(3+q^{-4}-2q^{-3}\lambda)(V^{+3}V^{0-}-V^{+0}V^{3-})\\
&  +\;(3+q^{-4}-2q^{-3})(V^{+-}V^{30}+V^{30}V^{+-})\nonumber\\
&  +\;q^{-6}(3q^{4}+1-2q\lambda)(V^{0-}V^{+3}-V^{3-}V^{+0})\nonumber\\
&  -\;q^{-2}\lambda(2q^{2}+2q\lambda+\lambda\lambda_{+})V^{+-}V^{+-},\nonumber
\end{align}
with $\eta_{\mu\nu}$ and $\varepsilon_{\mu\nu\rho\sigma}$ being q-analogs of
Minkowski metric and corresponding epsilon tensor (see also Appendix
\ref{AppA}). Specifying the two Casimirs for the different representations
finally yields the results:

\begin{enumerate}
\item[a)] (operator representation)
\begin{align}
C^{1}  &  =2q\lambda_{+}^{-2}(X\circ X)(\partial\circ\partial)+2q^{-2}%
\lambda_{+}^{-2}(X\circ\partial)(X\circ\partial)\\
&  +\;2q\lambda_{+}^{-1}X\circ\partial,\nonumber\\
C^{2}  &  =0,\nonumber
\end{align}

\item[b)] (spinor representation)
\begin{align}
C^{1}  &  =[[3]]_{q^{-4}}\lambda_{+}^{-2}\mbox{1 \kern-.59em {\rm l}}_{2\times
2},\\
C^{2}  &  =[[3]]_{q^{4}}(3+q^{4}+2q^{3}\lambda
)\mbox{1 \kern-.59em {\rm l}}_{2\times2},\nonumber
\end{align}

\item[c)] (vector representation)
\begin{align}
C^{1}  &  =2[[3]]_{q^{-4}}\lambda_{+}^{-2}%
\mbox{1 \kern-.59em {\rm l}}_{3\times3},\\
C^{2}  &  =0.\nonumber
\end{align}

\end{enumerate}

\section{Conclusion\label{Concl}}

Let us end with a few comments on our results. We dealt with quantum algebras
describing q-deformed versions of physical symmetries. This way we considered
$U_{q}(su_{2}),$ $U_{q}(so_{4}),$ and q-deformed Lorentz algebra. It was our
aim to provide a consistent framework of basic definitions and relations which
allow for representation theoretic investigations in physics.

In each case the starting point of our reasonings was the realization of
symmetry generators within q-deformed differential calculus. We listed the
relations of the corresponding symmetry algebras and presented expressions for
coproducts and antipodes on symmetry generators. We realized that the Hopf
structure of the symmetry generators allows us to define q-analogs of
classical commutators. Furthermore, we concerned ourselves with q-deformed
versions of such finite dimensional representations that play an important
role in physics, i.e. spinor and vector representation. With the help of these
representations we were able to write down q-deformed commutation relations
between symmetry generators and components of a spinor or vector operator.
Moreover, we calculated the adjoint action of the symmetry generators on each
other and obtained relations for quantum Lie algebras this way. Finally, we
gave expressions for the corresponding Casimir operators and specified them
for the different representations.

Our reasonings were in complete analogy to the classical situation, but
compared with the undeformed case our results are modified by terms
proportional to $\lambda=q-q^{-1}$. Hence, we regain in the classical limit as
$q\rightarrow1$ the familiar expressions. This observation nourishes the hope
that a field theory based on quantum group symmetries can be developed along
the same lines as its undeformed counterpart.

In this sense we also hope that implementing our results on a computer algebra
system like Mathematica will prove useful in a systematical search for new
q-identities. Furthermore, such an undertaking can be helpful to make it
obvious that everything presented in this article works fine.\vspace{0.4cm}

\noindent\textbf{Acknowledgements}\newline First of all we are very grateful
to Eberhard Zeidler for his invitation to the MPI Leipzig, his very
interesting and useful discussions, his special interest in our work and his
financial support. Also we want to express our gratitude to Julius Wess for
his efforts and his steady support. Furthermore we would like to thank Fabian
Bachmaier for teaching us Mathematica. Finally, we thank Dieter L\"{u}st for
kind hospitality.

\appendix

\section{q-Deformed quantum spaces\label{AppA}}

The aim of this appendix is the following. For quantum spaces of physical
importance we list their defining commutation relations as well as the
nonvanishing elements of their quantum metric and q-deformed epsilon tensor.

The coordinates $\theta^{\,\alpha},$ $\alpha=1,2,$ of two-dimensional
antisymmetrised quantum plane fulfill the relation \cite{Man88,SS90}
\begin{equation}
\theta^{1}\theta^{2}=-q^{-1}\theta^{2}\theta^{1},\quad q\in\mathbb{R},~q>1,
\end{equation}
whereas the spinor metric is given by a matrix $\varepsilon^{\alpha\beta}$
with nonvanishing elements
\begin{equation}
\varepsilon^{12}=q^{-1/2},\quad\varepsilon^{21}=-q^{1/2}.
\end{equation}
Furthermore, we can raise and lower indices as usual, i.e.
\begin{equation}
\theta^{\,\alpha}=\varepsilon^{\alpha\beta}\theta_{\beta},\quad\theta_{\alpha
}=\varepsilon_{\alpha\beta}\theta^{\beta},
\end{equation}
where $\varepsilon_{\alpha\beta}$ denotes the inverse of $\varepsilon
^{\alpha\beta}$.

In the case of three-dimensional q-deformed Euclidean space the commutation
relations between its coordinates $X^{A},$ $A\in\{+,3,-\},$ read
\begin{align}
X^{3}X^{\pm}  &  =q^{\pm2}X^{\pm}X^{3},\\
X^{-}X^{+}  &  =X^{+}X^{-}+\lambda X^{3}X^{3}.\nonumber
\end{align}
The nonvanishing elements of the quantum metric are
\begin{equation}
g^{+-}=-q,\quad g^{33}=1,\quad g^{-+}=-q^{-1}.
\end{equation}
Now, the covariant coordinates are given by
\begin{equation}
X_{A}=g_{AB}X^{B},
\end{equation}
with $g_{AB}$ being the inverse of $g^{AB}$. The nonvanishing components of
the three-dimensional q-deformed epsilon tensor take the form%
\begin{align}
\varepsilon_{-3+}  &  =1, & \varepsilon_{3-+}  &  =-q^{-2},\\
\varepsilon_{-+3}  &  =-q^{-2}, & \varepsilon_{+-3}  &  =q^{-2},\nonumber\\
\varepsilon_{3+-}  &  =q^{-2}, & \varepsilon_{+3-}  &  =-q^{-4},\nonumber\\
\varepsilon_{333}  &  =-q^{-2}\lambda. &  & \nonumber
\end{align}
In Sec. \ref{QuLie3dim} we need especially%
\begin{equation}
\epsilon_{AB}{}^{C}\equiv g^{CD}\varepsilon_{ABD},\quad\epsilon^{AB}{}%
_{C}\equiv g^{AD}g^{BE}\varepsilon_{DEC}.
\end{equation}

Next we come to four-dimensional Euclidean space. For its coordinates $X^{i},$
$i=1,\ldots,4,$ we have the relations
\begin{align}
X^{1}X^{j}  &  =qX^{j}X^{1},\quad j=1,2,\\
X^{j}X^{4}  &  =qX^{4}X^{j},\nonumber\\
X^{2}X^{3}  &  =X^{3}X^{2},\nonumber\\
X^{4}X^{1}  &  =X^{1}X^{4}+\lambda X^{2}X^{3}.\nonumber
\end{align}
The metric has the nonvanishing components
\begin{equation}
g^{14}=q^{-1},\quad g^{23}=g^{32}=1,\quad g^{41}=q.
\end{equation}
Its inverse denoted by $g_{ij}$ can again be used to introduce covariant
coordinates, i.e.
\begin{equation}
X_{i}=g_{ij}X^{j}.
\end{equation}
For the epsilon tensor of q-deformed Euclidean space with four dimensions one
can find as components%
\begin{align}
\varepsilon_{1234}  &  =q^{4}, & \varepsilon_{1432}  &  =-q^{2}, &
\varepsilon_{2413}  &  =-q^{2},\\
\varepsilon_{2134}  &  =-q^{3}, & \varepsilon_{4132}  &  =q^{2}, &
\varepsilon_{4213}  &  =q,\nonumber\\
\varepsilon_{1324}  &  =-q^{4}, & \varepsilon_{3412}  &  =q^{2}, &
\varepsilon_{2341}  &  =-q^{2},\nonumber\\
\varepsilon_{3124}  &  =q^{3}, & \varepsilon_{4312}  &  =-q, & \varepsilon
_{3241}  &  =q^{2},\nonumber\\
\varepsilon_{2314}  &  =q^{2}, & \varepsilon_{1243}  &  =-q^{3}, &
\varepsilon_{2431}  &  =q,\nonumber\\
\varepsilon_{3214}  &  =-q^{2}, & \varepsilon_{2143}  &  =q^{2}, &
\varepsilon_{4231}  &  =-1,\nonumber\\
\varepsilon_{1342}  &  =q^{3}, & \varepsilon_{1423}  &  =q^{2}, &
\varepsilon_{3421}  &  =-q,\nonumber\\
\varepsilon_{3142}  &  =-q^{2}, & \varepsilon_{4123}  &  =-q^{2}, &
\varepsilon_{4321}  &  =1.\nonumber
\end{align}
In addition to this there are the non-classical components%
\begin{equation}
\varepsilon_{3232}=-\varepsilon_{2323}=q^{2}\lambda.
\end{equation}

Now, we come to q-deformed Minkowski space \cite{SWZ91,OSWZ92,Maj91} (for
other deformations of spacetime and their related symmetries we refer to
\cite{Lu92,Cas93,Dov94,ChDe95,ChKu04,Koch04}). Its coordinates are subjected
to the relations
\begin{gather}
X^{\mu}X^{0}=X^{0}X^{\mu},\quad\mu\in{\{}0,+,-,3{\},}\\
X^{3}X^{\pm}-q^{\pm2}X^{\pm}X^{3}=-q\lambda X^{0}X^{\pm},\nonumber\\
X^{-}X^{+}-X^{+}X^{-}=\lambda(X^{3}X^{3}-X^{0}X^{3}),\nonumber
\end{gather}
and its metric is given by%
\begin{equation}
\eta^{00}=-1,\quad\eta^{33}=1,\quad\eta^{+-}=-q,\quad\eta^{-+}=-q^{-1}.
\end{equation}
As usual, the metric can be used to raise and lower indices. In analogy to the
classical case the q-deformed epsilon tensor has as nonvanishing components%
\begin{align}
\varepsilon_{+30-}  &  =-q^{-4}, & \varepsilon_{+-03}  &  =q^{-2}, &
\varepsilon_{3-+0}  &  =q^{-2},\\
\varepsilon_{3+0-}  &  =q^{-2}, & \varepsilon_{-+03}  &  =-q^{-2}, &
\varepsilon_{-3+0}  &  =-1,\nonumber\\
\varepsilon_{+03-}  &  =q^{-4}, & \varepsilon_{0-+3}  &  =-q^{-2}, &
\varepsilon_{30-+}  &  =q^{-2},\nonumber\\
\varepsilon_{0+3-}  &  =-q^{-4}, & \varepsilon_{-0+3}  &  =q^{-2}, &
\varepsilon_{03-+}  &  =-q^{-2},\nonumber\\
\varepsilon_{30+-}  &  =-q^{-2}, & \varepsilon_{+3-0}  &  =q^{-4}, &
\varepsilon_{3-0+}  &  =-q^{-2},\nonumber\\
\varepsilon_{03+-}  &  =q^{-2}, & \varepsilon_{3+-0}  &  =-q^{-2}, &
\varepsilon_{-30+}  &  =1,\nonumber\\
\varepsilon_{+0-3}  &  =-q^{-2}, & \varepsilon_{+-30}  &  =-q^{-2}, &
\varepsilon_{0-3+}  &  =1,\nonumber\\
\varepsilon_{0+-3}  &  =q^{-2}, & \varepsilon_{-+30}  &  =q^{-2}, &
\varepsilon_{-03+}  &  =-1,\nonumber
\end{align}
and%
\begin{align}
\varepsilon_{0-0+}  &  =q^{-1}\lambda, & \varepsilon_{-0+0}  &  =-q^{-1}%
\lambda,\\
\varepsilon_{0333}  &  =-q^{-2}\lambda, & \varepsilon_{3330}  &
=q^{-2}\lambda,\nonumber\\
\varepsilon_{3033}  &  =+q^{-2}\lambda, & \varepsilon_{3030}  &
=-q^{-2}\lambda,\nonumber\\
\varepsilon_{3303}  &  =-q^{-2}\lambda, & \varepsilon_{+0-0}  &
=-q^{-3}\lambda,\nonumber\\
\varepsilon_{0303}  &  =q^{-2}\lambda, & \varepsilon_{0+0-}  &  =q^{-3}%
\lambda.\nonumber
\end{align}

\end{document}